\documentclass[a4paper,11pt]{article}
\pdfoutput=1 % if your are submitting a pdflatex (i.e. if you have
\usepackage{jheppub} % for details on the use of the package, please
\usepackage[utf8]{inputenc}
\usepackage[T1]{fontenc} % if needed
\usepackage[dvipsnames]{xcolor}
\usepackage{multirow}
\usepackage{slashed}
\usepackage[normalem]{ulem}
\usepackage{sidecap}
\usepackage{mathtools}

\title{\boldmath Neutrino Portal to FIMP Dark Matter with an Early Matter Era}

\author[a]{Catarina Cosme}
\author[a]{, Ma\'ira Dutra}
\author[b]{, Teng Ma}
\author[a]{, Yongcheng Wu}
\author[c]{and Litao Yang}

\affiliation[a]{Ottawa-Carleton Institute for Physics,
Carleton University, 1125 Colonel By Drive, Ottawa, Ontario K1S 5B6, Canada}
\affiliation[b]{Physics Department, Technion - Israel Institute of Technology, Haifa, 3200003, Israel}
\affiliation[c]{Key Laboratory of Particle and Radiation Imaging (Ministry of Education) and Department of Engineering Physics, Tsinghua University, Beijing, 100084, China}

\emailAdd{ccosme@physics.carleton.ca}
\emailAdd{mdutra@physics.carleton.ca}
\emailAdd{t.ma@campus.technion.ac.il}
\emailAdd{ycwu@physics.carleton.ca}
\emailAdd{yanglt@mail.tsinghua.edu.cn}

\abstract{We study the freeze-in production of Feebly Interacting Massive Particle (FIMP) dark matter candidates through a neutrino portal. We consider a hidden sector comprised of a fermion and a complex scalar, with the lightest one regarded as a FIMP candidate. We implement the Type-I Seesaw mechanism for generating the masses of the Standard Model (SM) neutrinos and consider three heavy neutrinos, responsible for mediating the interactions between the hidden and the SM sectors. We assume that an early matter-dominated era (EMDE) took place for some period between inflation and Big Bang Nucleosynthesis, making the Universe to expand faster than in the standard radiation-dominated era. In this case, the hidden and SM sectors are easily decoupled and larger couplings between FIMPs and SM particles are needed from the relic density constraints. In this context, we discuss the dynamics of dark matter throughout the modified cosmic history, evaluate the relevant constraints of the model and discuss the consequences of the duration of the EMDE for the dark matter production. Finally, we show that if the heavy neutrinos are not part of the thermal bath, this scenario becomes testable through indirect detection searches. 
}

\usepackage{calligra}
\usepackage{calrsfs}
\DeclareMathAlphabet{\mathcalligra}{T1}{calligra}{m}{n}
\DeclareMathAlphabet{\pazocal}{OMS}{zplm}{m}{n}

\definecolor{darkpastelgreen}{rgb}{0.01, 0.75, 0.24}

\def\dm{\text{\tiny{DM}}}

\def\rh{\text{\tiny{RH}}}
\def\max{\text{\tiny{MAX}}}
\def\RD{\text{\tiny{RD}}}
\def\ERD{\text{\tiny{ERD}}}
\def\EMD{\text{\tiny{EMDE}}}
\def\EP{\text{\tiny{EP}}}

\def\M{\pazocal{M}}
\def\gs{\gamma}
\def\S{\pazocal{S}}
\def\la{\langle}
\def\ra{\rangle}
\def\to{\rightarrow}
\def\tob{\leftrightarrow}
\def\gev{\text{GeV}}

\def\df{\chi}  % dark fermion
\def\ds{S}     % dark scalar
\def\m{M}      % matter field

\def\lam{\lambda_\chi}  % dark fermion
\def\ss{\mathfrak{s}} % symbol for entropy subscription

\def\figureautorefname~#1\null{Fig.\,#1\null}
\def\tableautorefname~#1\null{Tab.\,#1\null}

\def\equationautorefname~#1\null{Eq.\,(#1)\null}

\begin{document} 
\maketitle
\flushbottom
\newpage
\section{Introduction}
\label{sec:intro}

The origin of dark matter (DM) is still one of the most important open problems in cosmology and particle physics. Dark matter is required to explain the galaxy rotation curves, the anisotropies of the Cosmic Microwave Background (CMB) and the observed structure of the Universe on large scales.  However, despite the large number of viable DM candidates (long-lived, cold, and sufficiently abundant), the nature of DM remains unknown (see Ref.~\cite{Bertone:2004pz} for a review).

Weakly Interacting Massive Particles (WIMPs) are the most popular DM candidates. These particles attained thermal equilibrium with the cosmic plasma in the early Universe and when their interactions with the Standard Model (SM) particles could no longer keep up against the expansion of the Universe, they decoupled from the thermal bath, yielding a frozen-out abundance. However, the lack of evidence of such DM candidates in detection experiments, the strong constraints on this framework~\cite{Arcadi:2017kky} and the absence of new particles at the LHC motivate exploring alternative scenarios. 

An interesting alternative to the WIMP paradigm is provided by the freeze-in mechanism~\cite{McDonald:2001vt, Hall:2009bx,Bernal:2017kxu}. According to this scenario, the DM abundance results from decays and annihilations of particles of the SM thermal bath and, due to small couplings between the dark and the visible sectors, the DM never reached chemical equilibrium with the cosmic bath. Since the interactions between DM and the SM particles are so feeble, this kind of DM candidates is called Feeble Interacting Massive Particles (FIMPs). The small couplings needed for the freeze-in allow to easily evade the stringent observational constraints, which makes the model appealing, although more difficult to test.

Another way to evade experimental constraints on DM is to assume a non-standard cosmological history of the Universe. Although we know that the Universe was radiation-dominated at the time of the Big Bang Nucleosynthesis (BBN), nothing prevents us to assume that another component dominated the Universe at early times, and it is interesting to study the consequences for the DM production. An early matter-dominated era (EMDE) would take place if some pressureless fluid - such as inflaton candidates~\cite{Allahverdi:2010xz}, meta-stable particles~\cite{Berlin:2016vnh,Tenkanen:2016jic,Berlin:2016gtr} and moduli fields~\cite{Vilenkin:1982wt,Starobinsky:1994bd,Dine:1995uk} - dominated the energy density of the universe prior BBN. In fact, when the EMDE is over, the dominant matter content might decay into SM degrees of freedom~\cite{Scherrer:1984fd}, reheating the visible sector and diluting the DM abundance. If DM was initially coupled to the thermal bath, the freeze-out needs to happen earlier than in the usual radiation-dominated era to overcome such dilution and agree with the relic density constraints. So, the interaction strengths of WIMPs to the thermal bath need to be weaker in the context of an EMDE. On the other hand, in order to overcome the dilution, FIMPs would only need to interact stronger to SM particles. This topic of DM production in models with a non-standard cosmology has gained increasing interest lately~\cite{Co:2015pka, Tenkanen:2016jic, Dror:2016rxc,DEramo:2017gpl, Hamdan:2017psw,Visinelli:2017qga,Drees:2017iod,Dror:2017gjq,DEramo:2017ecx,Bernal:2018ins,Bernal:2018kcw,Hardy:2018bph, Hambye:2018qjv,Biswas:2018iny,Fernandez:2018tfa,Chanda:2019xyl}.

In addition to DM, the neutrino sector constitutes another intriguing piece of our Universe. The discovery of neutrinos' mixing and masses, which are not included in the Standard Model (SM) of Particle Physics, is another reason to think of theories beyond the Standard Model (BSM). Thus, finding a scenario where these two phenomena are connected is an interesting possibility. In the literature, one can find several works that explore the neutrino portal dark matter, in the context of freeze-out~\cite{Macias:2015cna,Gonzalez-Macias:2016vxy, Batell:2017cmf,Berlin:2018ztp,Blennow:2019fhy} and freeze-in mechanisms~\cite{Chianese:2018dsz,Biswas:2016bfo,Abada:2014zra,Becker:2018rve, Chianese:2019epo, Kang:2010ha, Bian:2018mkl}. However, the study of a neutrino portal DM in a non-standard cosmology was only considered in the context of WIMPs~\cite{Berlin:2016gtr,Drees:2018dsj}.

In this sense, we will consider in this paper, for the first time, the freeze-in production of dark matter through the neutrino portal, when the Universe was dominated by a matter content at early times. The hidden sector of our model contains a fermion ($\chi$) and a complex scalar ($S$). Depending on the masses, either $\chi$ or $S$ can be a DM candidate with a $\mathbb{Z}_2$ symmetry ensuring the stabilization of the DM candidate. We implement the Type-I Seesaw mechanism for generating the masses of the SM neutrinos, considering three heavy neutrinos, $N$, which also mediate the visible and the hidden sectors. Therefore, the same couplings that provide neutrino's masses are also responsible for the DM phenomenology. The heavy neutrinos may or may not be coupled to the thermal bath, and we consider both possibilities separately in order to understand the impact of this hypothesis on our parameter space.

In terms of the cosmological evolution of the Universe, we assume that, after inflation, the inflaton decays but there is a matter component whose energy density is larger than the radiation energy density, taking over the evolution of the Universe. In this way, the Universe undergoes an early phase of radiation, followed by a matter-era domination until, at some temperature above the Big Bang Nucleosynthesis (BBN), this matter component decays and reheats the visible sector. We study the evolution of the Hubble parameter in each phase and the impact of this non-standard cosmology on the freeze-in production of the DM candidate.

In this paper, we present a discussion of the dynamics and phenomenology of this dark matter model, exploring the possibility of probing it through both direct and indirect detection. In regards to direct detection, we compare the spin-independent dark matter-nucleus scattering cross-section with bounds coming from Xenon-1T~\cite{Aprile:2017iyp} and projections of Xenon-nT~\cite{Aprile:2015uzo}. Indirect signals of WIMPs in the context of neutrino portal involve the internal bremsstrahlung~\cite{Garny:2013ama,Toma:2013bka,Giacchino:2013bta,Batell:2017rol} and cascade decays of mediators (see, for instance, \cite{Batell:2017rol}). Although challenging, indirect detection of FIMPs is a possibility that has been recently explored~\cite{Brdar:2017wgy,Heikinheimo:2018duk,Biswas:2019iqm}. As an interesting consequence of an EMDE, we explore the sensitivity of the indirect detection of \textit{FIMP annihilations} to heavy neutrinos, which subsequently decay into SM states. Our work is therefore a contribution to the recent and promising literature regarding the phenomenology of FIMP candidates, which involve constraints from direct~\cite{Battaglieri:2017aum,Hambye:2018dpi,Heeba:2019jho} and indirect~\cite{Brdar:2017wgy,Biswas:2019iqm} detection searches for dark matter, collider~\cite{Co:2015pka,Calibbi:2018fqf,Curtin:2018mvb,Belanger:2018sti,No:2019gvl} and accelerator~\cite{Heeba:2019jho} experiments and dark matter self-interaction~\cite{Hambye:2018dpi,Bernal:2018ins}. Frozen-in species are also subject to current cosmological bounds~\cite{Coffey:2020oir}.

This paper is organized as follows: in~\autoref{sec:model} we introduce the particle content of the model, while in~\autoref{sec:early} we parametrize the EMDE and discuss its impact on the Hubble rate. Then, in~\autoref{sec:freezein}, we discuss the freeze-in production of DM, as well as the theoretical constraints restricting our parameter space. We present the viable parameter space in ~\autoref{sec:results} and show the prospects for probing this scenario through direct and indirect detection in~\autoref{sec:pheno}. Finally, in~\autoref{sec:conclusions}, we present our conclusions. 

\section{The Model}
\label{sec:model}

In this work, we consider the Standard Model (SM) content and introduce an extra fermion, $\df$, and a scalar, $\ds$, both comprising the hidden/dark sector. The dark sector particles are odd under a dark $\mathbb{Z}_{2}$ symmetry, ensuring the stabilization of the fields, while the SM fields are even, and the dark matter candidate will be the lightest particle of the dark sector. In this model, $\ds$ and $\df$ are not in thermal equilibrium with each other and, for simplicity, we assume that the dark scalar $\ds$ does not acquire a vacuum expectation value (vev). In addition to this, the model contains three heavy neutrinos, $N$, responsible for mediating the interactions between the visible and the hidden sectors and generating the neutrino mass through Type-I Seesaw mechanism~\cite{Minkowski:1977sc,Yanagida:1980xy,Glashow:1980a,Mohapatra:1979ia,Schechter:1980gr}. The  Lagrangian of the model is as following:

\begin{equation}
\pazocal{L}=\pazocal{L}_{\rm SM}+\pazocal{L}_{\rm hidden}+\pazocal{L}_{\rm seesaw}+\pazocal{L}_{\rm portal},\label{total lagrangian}
\end{equation}
where the $\pazocal{L}_{\rm SM}$ corresponds to the SM Lagrangian, and the other terms are given by:
\begin{equation}
\pazocal{L}_{\rm hidden}= \overline{\df}(i\slashed{\partial}-m_\df)\df + |\partial_\mu \ds|^2 - m_\ds^2|\ds|^2+ V(\ds),
\label{hidden lagrangian}
\end{equation}

\begin{equation}
\pazocal{L}_{\rm seesaw}= \frac{1}{2} \overline{N}_\ell^i(i\slashed{\partial}\delta^{ij}-m_N^{ij})N_\ell^j - \left(\overline{L_L^i} Y_\nu^{ij} \tilde{H} (N_\ell^j)_R + h.c.\right),
\label{seesaw lagrangian}
\end{equation}

\begin{equation}
\pazocal{L}_{\rm portal}= - \left(\lambda_\chi^i \ds\overline{\df}(N_\ell^i)_R + h.c.\right),
\label{portal lagrangian}
\end{equation}
corresponding to the Lagrangian of the hidden sector, containing the kinetic and mass terms of $\df$ and $\ds$ (\autoref{hidden lagrangian}), the term responsible for the generation of neutrinos masses (\autoref{seesaw lagrangian}) and the term leading to the interactions between the hidden and the visible sectors (\autoref{portal lagrangian}). In this model, $Y_\nu^{ij}$ stands for the Yukawa coupling matrix (the structure will be discussed later), $L^i$ are the left-handed leptons, with $i=1,2,3$ being the generation index, the subscript $\ell$ denotes the interaction basis, $m_\chi$, $m_S$ and $m_N$ are the $\chi$, $S$ and $N$ masses, respectively, and  $\tilde{H} = i\sigma_2 H^*$, with $H$ being the SM Higgs doublet:
\begin{equation}
H=\begin{pmatrix}H^{+}\\
    H^0 
    \end{pmatrix}=\begin{pmatrix}G^{+}\\
\frac{v\,+h\,+\,iG_{0}}{\sqrt{2}} 
\end{pmatrix}, \label{Higgs doublet}
\end{equation}
where $v$ is the Higgs vev, $h$ is the physical Higgs, $G^\pm$ and $G^0$ are the Goldstone bosons from symmetry breaking and will become the longitudinal modes of $W^\pm$ and $Z$ respectively, and $H^\pm$ and $H^0$ are the charged and neutral components of the Higgs doublet before EWSB ($H^0$ is complex). In general, we can include $|S|^2|H|^2$ terms in the scalar potential. However, since we want to focus on the neutrino portal, such scalar portal interactions are temporarily ignored, and the exact form of the scalar potential $V(S)$ is, then, irrelevant.

\subsection{Type-I Seesaw}
\label{sec:model_seesaw}
After electroweak symmetry breaking (EWSB), the Lagrangian in~\autoref{seesaw lagrangian} provides masses and mixings for neutrinos through Type-I seesaw mechanism. The information from low energy neutrino measurements (neutrino mixings and masses/mass differences) is embedded in the interaction matrix $Y_{\nu}^{ij}$, which is parameterized in the Casas-Ibarra scheme~\cite{Casas:2001sr}:
\begin{align}
    \label{equ:ynu_ci}
    Y_{\nu} = \frac{i\sqrt{2}}{v}\,U_{\rm PMNS}\,m_\nu^{1/2}\,R\,m_N^{1/2},
\end{align}
where $U_{\rm PMNS}$ is the PMNS matrix containing three mixing angles ($\theta_{12},\theta_{23},\theta_{13}$) and three phases ($\delta_{\rm CP}, \alpha_1, \alpha_2$) and is parametrized as
\begin{align}
    U_{\rm PMNS} = \left(\begin{array}{ccc}
        1 & 0 & 0 \\
        0 & c_{23} & s_{23} \\
        0 & -s_{23} & c_{23}
    \end{array}\right)\cdot\left(\begin{array}{ccc}
        c_{13} & 0 & s_{13}e^{-i\delta_{\rm CP}} \\
        0 & 1 & 0 \\
        -s_{13}e^{i\delta_{\rm CP}} & 0 & c_{13} 
    \end{array}\right)\cdot \left(\begin{array}{ccc}
        c_{12} & s_{12} & 0 \\
        -s_{12} & c_{12} & 0 \\
        0 & 0 & 1
    \end{array}\right)\cdot \mathcal{P}
\end{align}
where $c_{ij}\equiv \cos\theta_{ij}$ and $s_{ij}\equiv\sin\theta_{ij}$, and $\mathcal{P}={\rm diag}(e^{i\alpha_1},e^{i\alpha_2},1)$. The value of these angles and phases are taken from the recent global fitting results~\cite{Esteban:2018azc}~\footnote{In our analysis, the two Majorana phases in $\mathcal{P}$ are not relevant and we will just set them to zero in further analysis.}. $m_{\nu/N}^{1/2}$ represent the diagonal matrices with square root of the eigen-masses ($\sqrt{m_{\nu/N}^i}$) in the diagonal entries and $R$ is an extra complex orthogonal matrix ($R^TR=\mathbb{I}$) parameterized by three complex angles. In fact, \autoref{equ:ynu_ci} is the generalization of the well-known seesaw formula $y\sim\frac{\sqrt{m_\nu m_N}}{v}$. In the three generations case, the complex orthogonal matrix $R$ is also important to determine the interaction patterns. If we consider large phases in the complex angles, the interaction can be highly enhanced while still keep neutrino masses light. However, in order to focus on the impact of a non-standard cosmological history, we will only consider the most trivial case where $R = \mathbb{I}$.

Before EWSB, there is no neutrino mixing, and the calculation is done directly with the interaction basis, whereas after EWSB the interaction in~\autoref{seesaw lagrangian} will induce the mixing among neutrinos:
\begin{align}
    \left(\begin{array}{c}
        (\nu_\ell^i)_L \\
        \widehat{(N_\ell^i)_R}
    \end{array}\right) = \mathbb{N}\left(\begin{array}{c}
        (\nu_m^i)_L \\
        \widehat{(N_m^i)_R}
    \end{array}\right),
\end{align}
where $\nu_\ell^i/N_\ell^i$ are in the interaction basis, $\nu_m^i/N_m^i$ are in the mass basis and $\widehat{\psi}\equiv\gamma_0C\psi^*$ is the Lorentz Covariant Conjugate (LCC) in the convention of~\cite{Pal:2010ih}.
The $6 \times 6$ mixing matrix $\mathbb{N}$ is parameterized by
\begin{align}
    \mathbb{N} \equiv \left(\begin{array}{cc}
        U & V\\
        X & Y
    \end{array}\right),
\end{align}
with, up to $\pazocal{O}\left(m_\nu/m_N\right)$:
\begin{align}
    U &\approx U_{\rm PMNS},\\
    V &\approx i\,U_{\rm PMNS}\,m_\nu^{1/2}\,R\,m_N^{-1/2},\\
    X &\approx i\,m_N^{-1/2}\,R^\dagger\, m_\nu^{1/2},\\
    Y &\approx \mathbb{I}_{3 \times 3}. 
\end{align}

\section{The Early Matter Era}
\label{sec:early}
\begin{figure}[!t]
    \centering
    \includegraphics[width=0.75\textwidth]{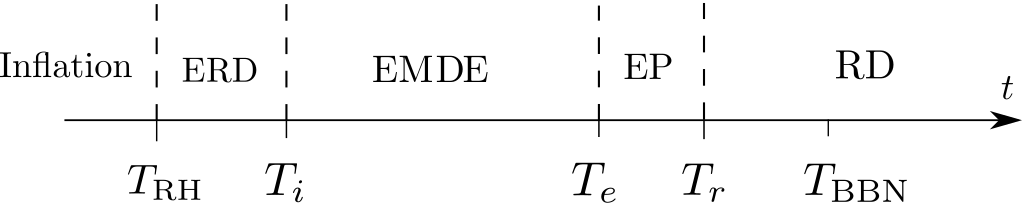}
    \caption{The history of the Universe with an early matter era.}
    \label{fig:history}
\end{figure}

An early matter era is a generic modification to the standard history of the Universe coming from many extensions of the standard model (SM) of particle physics. With such an era, the cosmic history can be simply described as in~\autoref{fig:history}. After inflation (which ends at the post-inflationary reheat temperature $T_\rh$), the ultra-relativistic SM species produced from inflaton decay dominate the total energy density and we have therefore a radiation-dominated (RD) era.

Let us assume that a BSM field $\m$ was once part of the thermal bath and decoupled while ultra-relativistic, so that its number density remains $n_M = \gamma_M \zeta(3)/\pi^2 T^3$ afterwards. In this case, the ratio between its energy density, $\rho_M$, and the energy density of the bath species, $\rho_R(T) = \pi^2/30 g_e(T) T^4 \propto a^{-4}$, dominated by radiation, would grow with the expansion of the Universe once $M$ becomes non-relativistic ($\rho_M \propto a^{-3}$):
\begin{equation}
\frac{\rho_M}{\rho_R} \rightarrow \frac{m_M n_M}{\rho_R} = \frac{45 \zeta(3)}{\pi^4} \frac{\gs_M}{g_e(T)} \frac{m_M}{T} \propto a\, m_M \,,
\end{equation}
where $\gs_M$ and $m_M$ are the internal degrees of freedom and the mass of the BSM field $M$, respectively, $g_e$ is the energetic degrees of freedom parameter, $a$ is the scale factor of the Universe and $T$ is the temperature of the SM thermal bath.

When $\rho_M(T)>\rho_R(T)$, at some temperature $T_i$, $M$ starts to dominate the cosmic expansion until some later time (at a temperature $T_e$), leading the so-called early matter-dominated era (EMDE). If $M$ is not to be a cosmic relic, it has to decay completely, and since it was once thermalized with the SM species, it should decay at least partially into the thermal bath. After $T_e$, the decay of $\m$ into radiation becomes efficient and makes $\rho_R$ to increase, producing entropy until $\m$ completely decays, at the so-defined reheat temperature $T_r$. This period is termed as entropy production (EP). After that, the Universe starts to be radiation-dominated again and evolves as in the usual case. Given our current understanding of the cosmic thermal history, the synthesis of light elements, the Big Bang nucleosynthesis (BBN), took place around the MeV scale in a radiation-dominated Universe, and therefore we must ensure that $T_r \gtrsim 4 \text{MeV} > T_{BBN}$~\cite{DeBernardis:2008zz,Hannestad:2004px, Kawasaki:2017bqm, Hasegawa:2019jsa}.

Since the expansion rate of the Universe depends on the total energy density it contains, the presence of $M$ could in principle affect the evolution of any species through the Hubble rate: 
\begin{equation}
H(a) = \frac{\sqrt{\rho_R(a)+\rho_M(a)}}{\sqrt{3}M_P} \,, 
\end{equation}
with $M_P \simeq 2.4 \times 10^{18}$\,GeV the reduced Planck mass.

For the purpose of studying the DM phenomenology, it is convenient to know how $H(a)$ depends on temperature. The temperature-scale factor relations during each period are different, as we review in what follows.

During a RD era, the Hubble parameter is the usually considered one:
\begin{align}\label{equ:Hubble_RD}
H_{\rm RD}(T) =    \frac{\pi\sqrt{g_e(T)}}{3\sqrt{10}}\frac{T^2}{M_P} \,.
\end{align}

During an isentropic EMDE, the energy density of the non-relativistic matter component $\m$ is:
\begin{align}\label{rhoM}
    \rho_{\m} &= \rho_{\m}^i\left(\frac{a^i}{a}\right)^3  = \frac{\rho_\m^i}{\mathfrak{s}^i} \mathfrak{s} = (m_\m Y_\m^i) \frac{2\pi^2}{45}g_{\mathfrak{s}}(T)T^3,
\end{align}
where $\mathfrak{s}$ is the entropy density, $g_\mathfrak{s}$ is the entropic degrees of freedom parameter, $Y_\m\equiv n_\m/\mathfrak{s}$ is the yield of the matter component, the variables with superscript $i$ are evaluated at the beginning of the EMDE, at $T_i$, and $Y_\m^i$ is the initial abundance of the matter component, which is a free parameter in our scenario. With $\rho_{\m}$ dominating the energy density, we have:
\begin{align}
    \label{equ:Hubble_EMDE}
    H_{\rm EMDE}(T) = \sqrt{\frac{\rho_\m}{3M_P^2}} = H_{\rm RD}(T_r)\sqrt{\frac{4}{3}\frac{m_\m Y_\m^i}{T_r}\frac{g_{\mathfrak{s}}(T)}{g_e(T_r)}}\left(\frac{T}{T_r}\right)^{3/2}.
\end{align}
From this expression, it is clear that for temperatures above $T_r$, $H_{\rm EMDE} > H_{\rm RD}$ and, consequently, the Universe expands faster.

The EP period is more involved, since in this case both $\rho_{\rm R}$ and $\rho_{\rm M}$ can change significantly. As long as $\m$ is much more abundant than the decay products, it is well-known that such an out-of-equilibrium decay can raise the total entropy of the bath species in a comoving volume $S = \mathfrak{s} a^3$~\cite{Scherrer:1984fd}: 
\begin{equation}\label{equ:entropy}
\frac{dS}{dt} = B_R \frac{\Gamma_M}{T} \rho_M a^3 \,, 
\end{equation}
with $B_R$ the branching ratio of the decay of $M$ into radiation. From the First Law of Thermodynamics in an expanding Universe, it follows:
\begin{equation}\label{equ:energydensities}
\frac{d\rho_i}{dt} + 3 H \rho_i (1+w_i) = \frac{T}{a^3} \frac{dS}{dt} 
\hspace{0.6cm} \Rightarrow \hspace{0.6cm}    
\begin{cases}
\frac{d\rho_M}{dt} + 3H \rho_M = - \rho_M \Gamma_M \\
\frac{d\rho_R}{dt} + 4H \rho_R = B_R \rho_M \Gamma_M \,,
\end{cases}
\end{equation} 
where $w_i$ is the ratio between the energy density and the pressure of a fluid $i$ ($w_M = 0$ for matter and $w_R = 1/3$ for radiation). 

\begin{figure}[!t]
\centering
\includegraphics[scale=0.64]{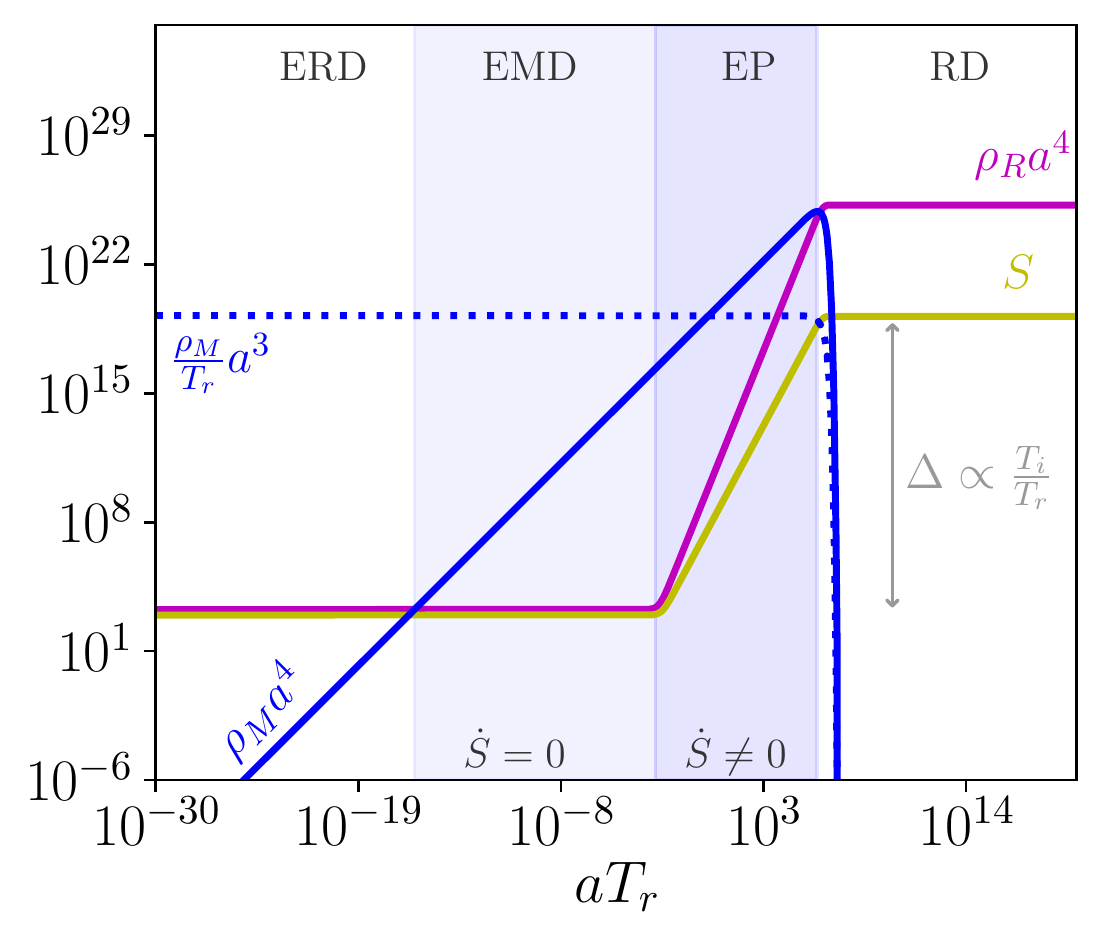}
\includegraphics[scale=0.64]{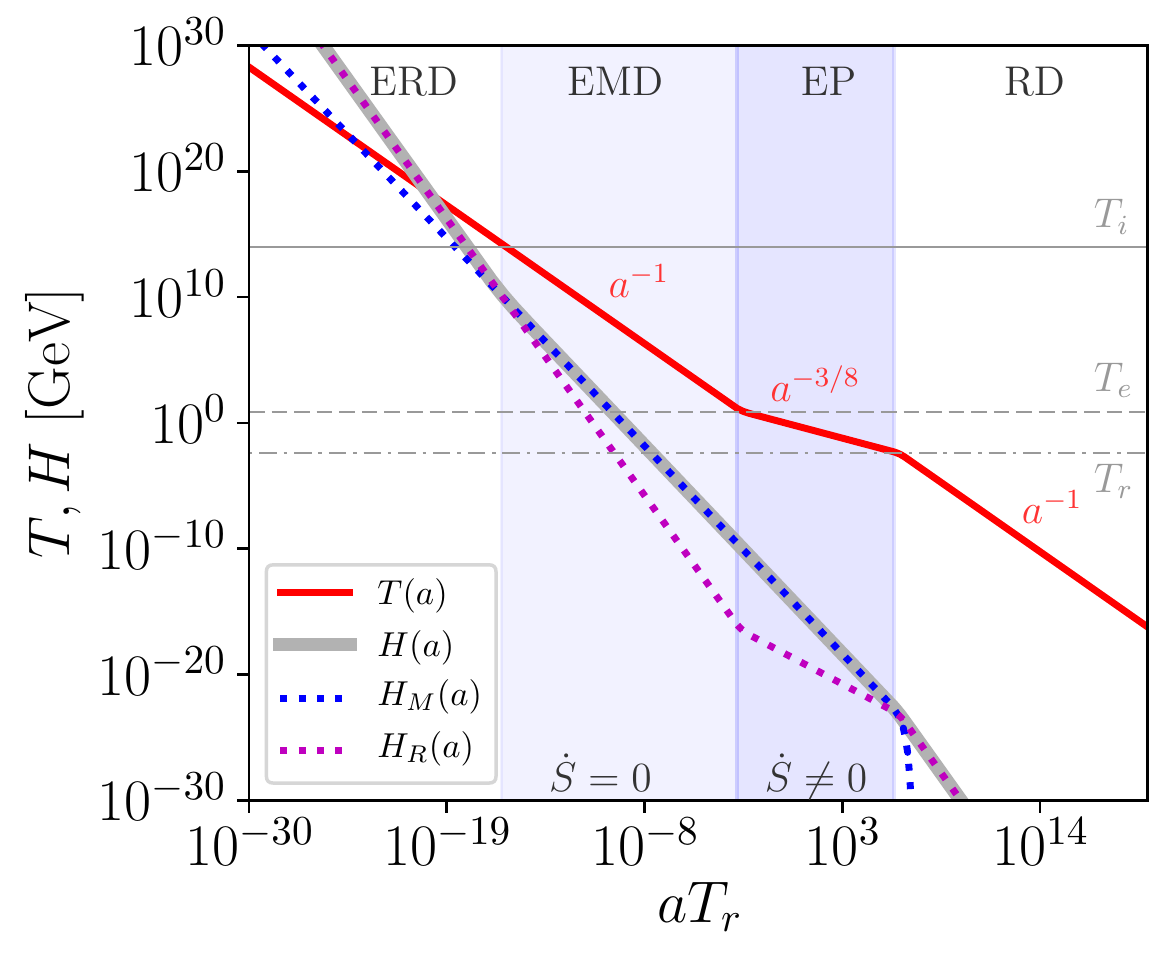}
\caption{\textit{Left:} coupled evolution of the matter (blue) and radiation (magenta) contents (see~\autoref{equ:energydensities}) and of the total entropy (yellow) (see~\autoref{equ:entropy}). \textit{Right:} consequent evolution of the temperature (red) and the Hubble rate (grey), along with the contributions of matter and radiation to the expansion (dotted blue and magenta). 
}
\label{Fig:thermalhistory}
\end{figure}

In~\autoref{Fig:thermalhistory}, we show the main physics at play in the non-standard thermal history we consider in this work, leaving a more detailed exposition to~\autoref{sec:freezein_evolution}, where the reader can also find the initial conditions we have used. In the left panel, we present the solutions of~\autoref{equ:entropy} and~\autoref{equ:energydensities} for the evolution of SM radiation (magenta) and matter (blue) contents, and of entropy (yellow), as functions of the scale factor normalized to the reheat temperature. When $\rho_M > \rho_R$, we have the early matter era, which might finish with a brief period of entropy production. We differentiate between the early radiation-dominated (ERD) era, before the EMDE, and the usual RD one since any relic produced by then might be diluted after the EP period. For simplicity, we assume that the usual radiation era follows just after the EP period, without any other dilution events. In the right panel, we present the evolution of the temperature of the SM bath $T(a)$ (red) and the Hubble rate $H(a)$ (grey). The dotted magenta and blue curves show the Hubble rate containing only radiation and matter, respectively. Details of this figure will be clear in what follows.

Although solving the coupled differential equations~\autoref{equ:entropy} and~\autoref{equ:energydensities} is needed for a precise description of the EP period, useful results can still be obtained with some reasonable approximations. 

Assuming that the decrease of $\rho_\m$ is negligible even during the EP ($\frac{d(\rho_\m a^3)}{da}\approx 0$) and that $\m$ still dominates the energy density, we have:
\begin{align}
    \frac{d(\rho_R a^4)}{da} \approx \sqrt{3}M_PB_R\Gamma_\m\left(\rho_\m^i (a^i)^3\right)^{1/2}a^{3/2}.
\end{align}
Then, the energy density of the radiation during the EP is solved to be:
\begin{align}
    \label{equ:rhoR_EP}
    \rho_R \approx \frac{2}{5}\sqrt{3}M_PB_R\Gamma_\m\left(\rho_\m^i (a^i)^3\right)^{1/2}a^{-3/2}.
\end{align}

On the other hand, temperature is defined through the energy density of radiation, $\rho(T) \propto T^4$. Thus, instead of the usual relation $T\sim a^{-1}$, valid for an isentropic expansion, we have $T\sim a^{-3/8}$ while entropy is being produced (see the right panel of~\autoref{Fig:thermalhistory}). Finally, the Hubble parameter during the EP is:
\begin{align}
    H_{\rm EP}(T) \approx \sqrt{\frac{\rho_M}{3M_P^2}} = \sqrt{\frac{\rho_\m^i (a^i)^3}{3M_P^2a^3}}= \frac{5}{2}\frac{1}{3M_P^2B_R\Gamma_\m}\frac{\pi^2}{30}g_e(T)T^4 \,.
\end{align}

We can further simplify this expression by identifying $T_r$ as the temperature at which the decay width $\Gamma_\m$ of $M$ is comparable to the Hubble parameter: $\Gamma_\m = \kappa H_{\rm RD}(T_r)$. In terms of the dimensionless variable $\kappa$, we have:
\begin{align}
    \label{equ:Hubble_EP}
    H_{\rm EP}(T) = H_{\rm RD}(T_r)\frac{5}{2} \frac{1}{\kappa B_R} \frac{g_e(T)}{g_e(T_r)}\left(\frac{T}{T_r}\right)^4.
\end{align}

As a consequence of the entropy production, the number densities will be diluted. Hence, we would like to identify a variable that can quantify the dilution. From~\autoref{equ:entropy}, we have:
\begin{align}
    \label{equ:entropyODE}
    S^{1/3}dS &= \sqrt{3}\left(\frac{2\pi^2}{45}g_{\mathfrak{s}}\right)^{1/3}B_RM_P\Gamma_\m\rho_\m^{1/2}a^3da.
\end{align} 

Integrating~\autoref{equ:entropyODE} for a sufficiently long EP period and assuming $\rho_R (T_r) \sim \rho_M (T_r) \approx \rho_M(T_i) (a_i/a_r)^3$, we can use~\autoref{rhoM} to find:
\begin{equation}
    \frac{S_r}{S_i} \approx \frac{4}{15} \sqrt{\frac{2\pi}{3}} B_R^{3/4} \kappa^{5/4} \bar g_s^{1/4} \frac{m_M Y_M^i}{\sqrt{M_P \Gamma_M}}\,.
\end{equation}
From the equation above, we can see that the matter component needs to be sufficiently long lived for our purposes. By further using $\Gamma_M = \kappa H_{RD}(T_r)$, we have:
\begin{equation}\label{deltadef}
\frac{S_r}{S_i} \approx \frac{4}{3} \left(\frac{2}{5}\kappa B_R\right)^{3/4} \frac{m_M^i Y_M^i}{T_r}.
\end{equation}

Thus, we define a dilution factor $\Delta$ as:
\begin{align}
    \Delta \equiv \frac{m_\m Y_\m^i}{T_r} \,.
\end{align}

Notice that the continuity of the Hubble rate among different periods also provides useful qualitative relations. Requiring:
\begin{align}
    H_{\rm RD}(T_i) &= H_{\rm EMDE}(T_i)\nonumber\\
    H_{\rm EMDE}(T_e) &= H_{\rm EP}(T_e) \nonumber \\
    H_{\rm EP}(T_r) &= H_{\rm RD}(T_r) \,,
\end{align}
\autoref{equ:Hubble_RD}, \autoref{equ:Hubble_EMDE} and~\autoref{equ:Hubble_EP} lead to:
\begin{align}
    \label{equ:EMDE_relations}
    \Delta = \frac{3}{4}\frac{g_e(T_i)}{g_{\mathfrak{s}}(T_i)}\frac{T_i}{T_r},\quad \frac{T_e}{T_r} = \left(\Delta \frac{4}{3}\frac{g_{\mathfrak{s}}(T_e)g_e(T_r)}{g_e^2(T_e)}\right)^{1/5}, \quad \kappa B_R = \frac{5}{2}.
\end{align}

With these relations, we can fully determine the cosmic history with an EMDE by only specifying $T_r$ and $T_i$, with the properties of the matter component $\m$ embedded in the latter. In summary, the Hubble rates that we will use in the DM relic density calculation for different periods are:
\begin{align}
    \label{equ:Hubble_Final_ERD}
    &{\rm ERD:}& H_{\rm ERD}(T) & = \frac{\pi\sqrt{g_e(T)}}{3\sqrt{10}}\frac{T^2}{M_P}\\
    \label{equ:Hubble_Final_EMDE}
    &{\rm EMDE:}& H_{\rm EMDE}(T) &= H_{\rm RD}(T_r)\sqrt{\frac{4}{3}\Delta\frac{g_{\mathfrak{s}}(T)}{g_e(T_r)}}\left(\frac{T}{T_r}\right)^{3/2}\\
    \label{equ:Hubble_Final_EP}
    &{\rm EP:}&  H_{\rm EP}(T) &= H_{\rm RD}(T_r)\frac{g_e(T)}{g_e(T_r)}\left(\frac{T}{T_r}\right)^4\\
    \label{equ:Hubble_Final_RD}
    &{\rm RD:}&  H_{\rm RD}(T) & = \frac{\pi\sqrt{g_e(T)}}{3\sqrt{10}}\frac{T^2}{M_P}
\end{align}
where $\Delta$ and $T_e$ are determined from~\autoref{equ:EMDE_relations}.

\section{Freeze-in Production in an Early Matter Era}
\label{sec:freezein}

In the freeze-in mechanism~\cite{McDonald:2001vt,Hall:2009bx}, the DM interaction strength with the thermal bath fields is so weak that DM cannot reach thermal equilibrium with them. Therefore, if DM particles were not initially present in the SM thermal bath, which might also contain BSM fields, the bath species could be able to produce DM without back reaction. The freeze-in of such FIMP DM candidates would happen whenever kinematically possible: while thermal bath particles are abundant enough (for temperatures above their Boltzmann suppression) and have enough energy to produce FIMPs (for temperatures above DM Boltzmann suppression). As opposed to the freeze-out production of WIMPs, the FIMP relic abundance is \textit{proportional} to the annihilation cross section or the partial decay width into DM. If we consider the simplest case where a dark matter particle $\chi$ is produced through a heavy resonance $N$ decay channel, $N \to \chi S$, with $N$ coupled to the SM thermal bath, the DM relic abundance from this decay is generically approximated as:
\begin{align}
\Omega_\chi h^2 \propto \frac{m_\chi \Gamma_{N \to \chi S}}{m_N^2},
\end{align}
where $m_{\chi}$ and $m_N$ are the masses of $\chi$ and $N$  and $\Gamma_{N \to \chi S}$ is the partial decay width of $N$ into $\chi$ and $S$.

Particles which are not coupled to the thermal bath might be much less abundant than the SM species. In this work, we consider both $\chi$ and $S$ as FIMP candidates. The mediator $N$ between the FIMPs and the SM fields (Higgs bosons and leptons) is considered separately as thermalized and non-thermalized with the SM bath. As it will become clear in~\autoref{sec:pheno}, this is a crucial point for the phenomenology of our scenario. 

In what follows, we present the processes contributing to the production of our FIMP candidates and we give an approximate expression of the relic density of a generic FIMP candidate taking into account an early matter-dominated era. We then discuss the conditions for the heavy neutrinos to be thermalized and the consistency conditions for the freeze-in mechanism to hold.

\subsection{Reaction Rate Densities}
\label{sec:freezein_processes}

The evolution of dark matter is governed by the Boltzmann fluid equation for its total number $N_\dm = n_\dm a^3$:
\begin{equation}\label{dNdt}
\frac{dN_\dm}{dt} = (\dot n_\dm + 3H(t)n_\dm) a^3 = R_\dm (t) a^3 \,,
\end{equation}
where $R_\dm(t)$ is the reaction rate density, accounting for all the production and loss of dark matter particles in a comoving volume $a^3$. Since we are concentrated here in the freeze-in mechanism, we hereafter regard $R_\dm(T)$ as just production rate densities.

\begin{figure}[!t]
\centering
\includegraphics[width=\textwidth]{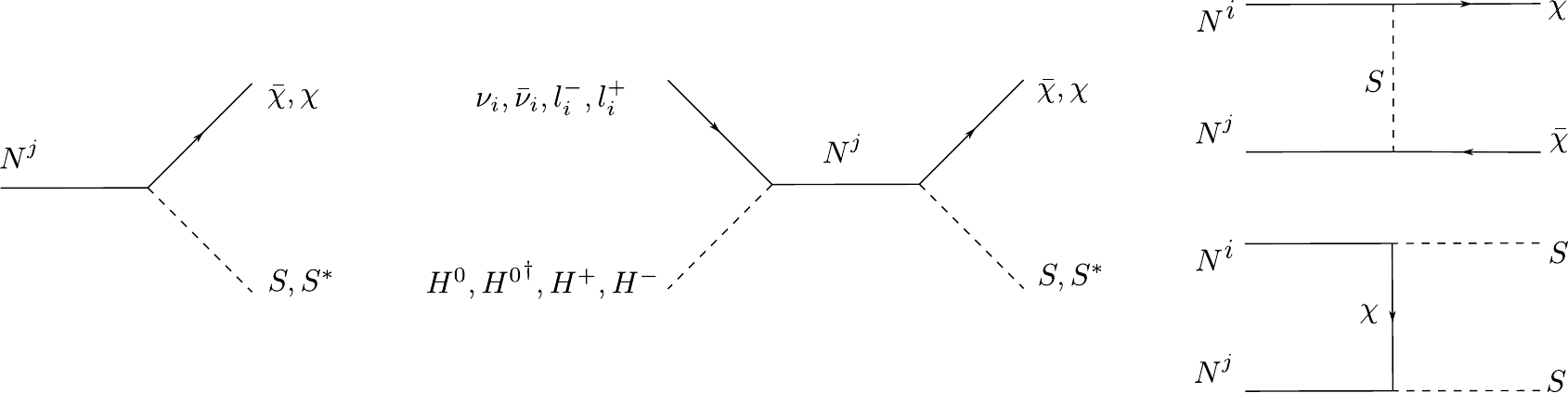}
\caption{Processes contributing to the freeze-in production of our dark matter candidate, considered as either $\chi$ or $S$. 
}
\label{Fig:diagrams}
\end{figure}

In this work, we are going to explore both dark scalar ($S$) and dark fermion ($\chi$) as dark matter candidates. Whenever kinematically allowed, the channels contributing for their production before EWSB, depicted in~\autoref{Fig:diagrams}, are:
\begin{equation}\label{Rew}
R_{\chi/S}(T)\Big|_{EW} = R_{\chi/S}^{N \to \bar\chi S}(T) + R_{\chi/S}^{N N \to \bar\chi \chi/S^* S}(T) + R_{\chi/S}^{\nu_i H^0 \to \bar\chi S}(T) + R_{\chi/S}^{l_i^- H^+\to \bar\chi S}(T) \,. 
\end{equation}
After EWSB, we have:
\begin{equation}\label{Rewb}
\begin{split}
R_{\chi/S}(T)\Big|_{\slashed{EW}} = & R_{\chi/S}^{N \to \bar\chi S}(T) 
+ R_{\chi/S}^{\nu_i \to \bar\chi S}(T) 
+ R_{\chi/S}^{N N \to \bar\chi \chi/S^* S}(T) + R_{\chi/S}^{\nu_i \nu_i \to \bar\chi \chi/S^* S}(T)\\
& + R_{\chi/S}^{\nu_i h \to \bar\chi S}(T) + R_{\chi/S}^{\nu_i Z \to \bar\chi S}(T) + R_{\chi/S}^{l_i^- W^+\to \bar\chi S}(T) \,.
\end{split}
\end{equation}
Notice that we are neglecting the contributions of the processes $S \to N \chi$ and $\chi \to N S$ for the productions of $\chi$ and $S$ respectively since they are non-thermal and their initial densities are negligible ($n_\chi, n_S \ll n_{N}, n_{\nu_i}, n_{H}$).

For a process $1 \to 23$, in which species $1$ is thermalized with other species than $2$ and $3$, the rate density for the production of species $2$ is approximately given by
\begin{equation}
R_2^{1\to 23} \approx n_1 \Gamma_{1\to 23}\,,
\label{rate density decay}
\end{equation}
where $\Gamma_{1\to 23}$ is the decay width of species $1$ into species $2, 3$. It is a good approximation to consider the number density of the decaying field $n_1$ as a Maxwell-Boltzmann distribution:
\begin{equation}
n_1 (T) \approx \frac{\gs_1}{2\pi^2} m_1^2 T K_2\left(\frac{m_1}{T}\right)\,,
\end{equation}
where $\gs_1$ is the internal degrees of freedom of species $1$ and $K_i$ is the modified Bessel function of second kind and order $i$.

In the case of the decay $N \to \bar \chi S$, the rate density is thus given by:
\begin{equation}\begin{split}
R_{\chi/S}^{N \to \bar\chi S} &= \frac{3|\lam|^2}{16\pi^3}m_N^3 T K_2\left(\frac{m_N}{T}\right) (1+r_\chi^2-r_S^2)\sqrt{1-(r_\chi+r_S)^2}\sqrt{1-(r_\chi-r_S)^2}  \\
& \approx \frac{3|\lam|^2}{16\pi^3}m_N^3 T K_2\left(\frac{m_N}{T}\right) (1-\epsilon^2)^2\,,
\label{Rdecay}
\end{split}\end{equation}
where, for convenience, we define the dimensionless parameters $r_i \equiv m_i/m_N$ and $\epsilon \equiv r_S/(1-r_\chi)$, with $0<\epsilon<1$.

For a process $12 \to 34$, in which species $1$ and $2$ are thermalized between themselves, the rate density for the production of species $3$ is given by:
\begin{equation}\begin{split}
R_3^{12\to 34} &\equiv n_1^{eq} n_2^{eq} \la \sigma v \ra_{12\to 34} \\
&= \frac{\S_{12}\S_{34}}{32(2\pi)^6}\int ds \frac{\sqrt{\lambda(s,m_3^2,m_4^2)}}{s} \int d\Omega_{13} |\M|^2 
\int_{m_1}^\infty dE_1 \int_{E_2^-}^{E_2^+} dE_2 f_1^{eq} f_2^{eq}\\
&\approx \frac{\S_{12}\S_{34}}{32(2\pi)^6}T \int ds \frac{\sqrt{\lambda(s,m_3^2,m_4^2)}}{s} \int d\Omega_{13} |\M|^2 \Big(\frac{\sqrt{\lambda(s,m_1^2,m_2^2)}}{\sqrt{s}}  K_1\left(\frac{\sqrt{s}}{T}\right) + \\
&\hspace{8cm} + |m_1^2-m_2^2| e^{-\sqrt{s}/T} \frac{\sqrt{s}+T}{s} \Big) \,,
\end{split}
\label{rate density scatt}
\end{equation}
where the approximation holds for initial states with Maxwell-Boltzmann statistics at zero chemical potential. The symmetrization factor $\S_{A(B)}=1/N_{A(B)}!$ accounts for $N_A ~(N_B)$ identical particles in the initial (final) state, $\lambda(x,y,z) = (x-(\sqrt{y}+\sqrt{z})^2)(x-(\sqrt{y}-\sqrt{z})^2)$ is the K\"allen function, $|\M|^2 $ is the squared amplitude of the process, $E_i$ is the energy of the particle $i$, $f_i^{eq}$ is the phase space equilibrium distribution function of the particle $i$ and $s$ is the center-of-mass energy squared. 

The integrated squared amplitudes of the processes of the kind $l_i H \to \chi S$, with an s-channel exchange of $N$, are all given by:
\begin{equation}
\int d\Omega_{13} |\M|^2 = 4\pi |\lam|^2\sum_j|Y_\nu^{ij}|^2 \left(1+\frac{m_N^2}{s}\right) \frac{s^2 \left(1+(m_l^2-m^2_H(T))/s\right)\left(1+(m^2_\chi-m_S^2)/s\right)}{(s-m_N^2)^2+m_N^2\Gamma_N^2}\,.    
\end{equation}

When the initial states from the thermal bath have enough energy to produce $N$ on-shell, we can use the narrow width approximation (NWA) to have an approximate expression for the rate in the resonant region:
\begin{equation}\label{equ:NWA}
\begin{split}
R_\chi^{\text{\ensuremath{l_{i}}H\ensuremath{\rightarrow}\ensuremath{\ensuremath{\chi}}S}}(T)\Big|_{NWA} = &\frac{|\lambda_{\chi}|^{2}\sum_{j}|Y_{\nu}^{ij}|^{2}}{16\left(2\pi\right)^{4}} \frac{T}{\Gamma_N(T)} m_N^4 \Theta(m_N-\text{max}(m_H(T)+m_l,m_\chi+m_S))\\
& \times \sqrt{\lambda(1,r_\chi^2,r_S^2)} (1+r_\chi^2-r_S^2) (1+r_l^2-r_H^2(T))\\
& \times \left[
\sqrt{\lambda(1,r_l^2,r_H^2(T))}K_1\left(\frac{m_N}{T}\right)+\left(1+\frac{T}{m_N}\right)|r_l^2-r_H^2(T)|e^{-\frac{m_H(T)}{T}}\right]\,.
\end{split}
\end{equation}

Far from the resonance, the s-channel processes contribute with the following production rate:
\begin{equation}\label{rateschannel}
R_\chi^{l_i H \to \chi S} \approx \frac{16|\lam|^2\sum_j|Y_\nu^{ij}|^2}{3(2\pi)^5}\times
\begin{cases}
T^4 I(2,c_h,x_\chi,x_S), & \text{for}~ T \gg m_N \\
\frac{T^6}{m_N^2} I(4,c_h,x_\chi,x_S), & \text{for}~ T \ll m_N \,,
\end{cases}
\end{equation}
where the only approximation is $m_l \ll m_H(T)$. Here, we take into account the thermal corrections for the Higgs mass according to the high-temperature expansion approximation of the finite temperature scalar potential~\cite{Espinosa:2011ax}. The mass parameter is $\mu^2(T) = -m_h^2/2 + c_h T^2$, where $c_h = \frac{1}{16}(3 g^2+g'^2)+\frac{1}{4}y_t^2 + \frac{1}{2}\lambda$, $g$ and $g^\prime$ are the electroweak $SU(2)_L$ and hypercharge $U(1)_Y$ gauge couplings, $y_t$ is the top Yukawa coupling and $\lambda$ is the Higgs quartic self-coupling. Before EWSB, $H^0$ and $H^\pm$ have the same mass which, at high temperature, is approximately $\sqrt{c_h}T$, whereas after EWSB, the Higgs develops a vev which varies with temperature and $G^\pm$ and $G^0$ become the longitudinal modes of $W^\pm$ and $Z$, respectively. Then, $h$, $W^\pm$ and $Z$ masses evolve with the vev accordingly. We have defined the following integral:
\begin{equation}
I(n,c_h,x_\chi,x_S) \equiv \int dz \, z^n \left(1-\frac{c_h}{z^2}\right)\left(1+\frac{x_\chi^2-x_S^2}{z^2}\right)\left(\left(1-\frac{c_h}{z^2}\right)K_1(z)+\frac{c_h}{z^2}\left(1+\frac{1}{z}\right)e^{-z}\right) \,,
\end{equation}
with $x_i \equiv m_i/T$ and $z \equiv \sqrt{s}/T \gtrsim \text{max}(\sqrt{c_h},x_\chi+x_S)$, accounting for a numerical factor of order one for the most of our parameter space.

In the case of the t-channel, in the limit $s \gg {\rm max}(4m_N^2, 4m_\chi^2)$ or $s \gg {\rm max}(4m_N^2, 4m_S^2)$, the contribution for the production rate of $\chi$ or $S$ reads:
\begin{equation}\label{ratetchannel}
R_{\chi/S}^{N N \to \bar \chi \chi / S^* S} \approx \frac{9|\lam|^4}{4(2\pi)^5} T^4 I_{S/\chi}(x_{S/\chi},\text{max}(x_N,x_{\chi/S})) \,, 
\end{equation}
where
\begin{equation}
I_S(x_S,\text{max}(x_N,x_\chi)) \equiv \int_{\text{max}(x_N,x_{\chi})} dz \, z^2 K_1(z) \left(1-\frac{1}{2(1+x_S^2/z^2)} - \frac{x_S^2}{z^2} \log\left(1+\frac{z^2}{x_S^2}\right) \right),
\end{equation}
and
\begin{equation}
I_\chi(x_\chi,\text{max}(x_N,x_S)) \equiv \int_{\text{max}(x_N,x_{S})} dz \, z^2 K_1(z) \left(-1+\frac{1}{2}\left(1+\frac{2x_\chi^2}{z^2}\right)\log\left(1+\frac{z^2}{x_\chi^2}\right) \right) \,.
\end{equation}

Since neutrinos are the only ultra-relativistic species in the freeze-in processes, and their Fermi-Dirac statistics do not provide a significant suppression in the rates, we safely use the Maxwell-Boltzmann approximation in our numerical code.

\subsection{Contributions to FIMP Relic Density}
\label{sec:freezein_relic}

In terms of the dark matter yield $Y_\dm \equiv N_\dm/S = n_\dm/\mathfrak{s}$, the evolution of the dark matter abundance given in~\autoref{dNdt} reads, in general:
\begin{equation}\label{equ:DMevol}
    \frac{dY_\dm}{da} = \frac{R_\dm(Y_\dm,a)}{a \mathfrak{s}(a)H(a)} - \frac{Y_\dm}{S}\frac{dS}{da},
\end{equation}
The second term in equation above accounts for the dilution in the dark matter abundance due to the entropy production which might take place at the end of an early matter era. In this case, this equation becomes coupled to the set of~\autoref{equ:entropy} and~\autoref{equ:energydensities}.

As soon as dark matter is produced, its number of particles in a comoving volume, $N_\dm$, becomes constant. Assuming no other entropy production period for temperatures below $T_r$, the yield of DM, and therefore its relic density, becomes also constant. 

By definition, the relic density of dark matter today reads:
\begin{equation}
\Omega_\dm^0 h^2 \simeq \frac{m_\dm}{\rm GeV} \frac{Y_\dm^0}{3.60 \times 10^{-9}}\,,   \label{relic dens}
\end{equation}
with $Y_\dm^0 = Y_\dm (T_0)$ and $T_0$ the present CMB temperature.

In the case of the FIMP dark matter candidate, the reaction rate density $R_\dm$ contains only its production term and does not depend on the dark matter yield itself, as there is no back reaction into the thermal bath. A fair approximate expression for the relic density today, taking into account the early matter era, can be therefore found from the results of~\autoref{sec:early}, as we describe in what follows.

In the presence of an EMDE, it is crucial to determine the DM yield accumulated before reheating:
\begin{equation}
Y_\dm^0 = Y_\dm (T_r) + \frac{135 \sqrt{5}}{\pi^3\sqrt{2}} M_P \int_{T_0}^{T_r} dT \frac{g_\ss^*(T)}{g_\ss(T) \sqrt{g_e(T)}} \frac{R_\dm(T)}{T^6} \,,
\end{equation}
where $g_\ss^* \equiv 1 + \frac{1}{3}\frac{d \ln g_\ss}{d \ln T}$ and $Y_\dm(T_r)$ is the yield of DM at $T_r$.

Under entropy production (from $T_e$ to $T_r$), we can define a different comoving yield given by $\tilde Y_\dm \equiv \frac{N_\dm}{\Phi} = \frac{n_\dm}{\Phi a^{-3}}$, with the nearly constant dimensionless quantity $\Phi = a^3 \rho_M/T_r$~\cite{Giudice:2000ex}. From~\autoref{equ:rhoR_EP}, we can see that:
\begin{equation}
\Phi a^{-3} (T) \approx \frac{\pi^2}{30} \frac{g_e^2(T)}{g_e(T_r)} \frac{T^8}{T_r^5},    
\end{equation}
and therefore its relations with $Y_\chi$ at $T_r$ and $T_e$ read: 
\begin{align*}
\tilde Y_\dm (T_r) &= Y_\dm (T_r) \frac{\ss(T_r)}{\Phi a^{-3}(T_r)} = Y_\dm (T_r) \frac{4}{3} \frac{g_\ss (T_r)}{g_e(T_r)}, \\
\tilde Y_\dm (T_e) &= Y_\dm (T_e) \frac{\ss(T_e)}{\Phi a^{-3}(T_e)} = Y_\dm (T_e)\frac{1}{\Delta}.
\end{align*}

During an EP period, the evolution of $\tilde Y_\dm$ is given by~\cite{Dutra:2019gqz}:
\begin{equation}
\frac{d \tilde Y_\dm}{dT} = - \frac{8}{3} g_e^*(T) \frac{R_\dm (T)}{H_{EP}(T) T \Phi a^{-3}(T)}\,,
\end{equation}
where $g_e^* \equiv 1 - \frac{1}{4}\frac{d \ln g_e}{d \ln T}$.

The yield of dark matter today has therefore the following contributions:
\begin{equation}\label{equ:Y0}
Y_\dm^0 = y_\RD + \frac{3}{4}\frac{g_e(T_r)}{g_\ss(T_r)} \left[ y_\EP + \frac{1}{\Delta}  (y_\EMD + y_\ERD) \right]\,, 
\end{equation}
with
\begin{equation*}
\begin{split}
y_\RD &\equiv \frac{135 \sqrt{5}}{\pi^3 \sqrt{2}} M_P \int_{T_0}^{T_r} dT \frac{g_s^*(T)}{g_s(T)\sqrt{g_e(T)}} \frac{R_\dm (T)}{T^6} \\
y_\EMD &\equiv \frac{135 \sqrt{15}}{2\pi^3 \sqrt{2}} \frac{1}{\sqrt{\Delta T_r}} M_P \int_{T_e}^{T_i} dT \frac{g_\ss^*(T)}{g_\ss^{3/2}(T)} \frac{R_\dm (T)}{T^{11/2}} \\
y_\EP &\equiv \frac{240 \sqrt{10}}{\pi^3} g_e^{3/2}(T_r) M_P T_r^7 \int_{T_r}^{T_e} dT \frac{g_e^*(T)}{g_e^3(T)} \frac{R_\dm (T)}{T^{13}} \\
y_\ERD &\equiv \frac{135 \sqrt{5}}{\pi^3 \sqrt{2}} M_P \int_{T_i}^{T_\rh} dT \frac{g_\ss^*(T)}{g_\ss(T)\sqrt{g_e(T)}} \frac{R_\dm (T)}{T^6}\,.
\end{split}
\end{equation*}
Notice that all information regarding the specific FIMP model is encoded in the production rate density $R_\dm(T)$\footnote{In this work, all the multidimensional integral are calculated numerically using the {\tt CUBA} library~\cite{Hahn:2004fe}.}.

From the expressions above we can extract important information regarding the nature of the freeze-in. If the dominant contribution to the production rate depends on temperature through some power law $T^n$, we can predict whether the freeze-in would happen at the highest or at the lowest scale available of a given cosmological era~\cite{Dutra:2019xet,Bernal:2019mhf}. The kind of freeze-in is therefore referred to as infrared (IR) or ultraviolet (UV) with respect to that era. In the case of a $1\to 2$ or a resonant $2\to 2$ process, the freeze-in happens at temperatures close to the decaying field or mediator mass. On the other hand, any production channel is only possible for temperatures above the Boltzmann suppression of the heaviest particle involved. In summary, the freeze-in temperature $T_{FI}$ can be determined as follows:
\begin{equation}
\begin{aligned} \label{equ:FItemperature}
&T_{FI} \big|^{X \,\text{on-shell}} \sim m_X \\
&T_{FI} \big|^{\text{ERD/RD}} \sim \begin{cases}
\text{max}(T_{i/0},m_\chi,m_S,m_N), &n<5 \\
\text{max}(T_{\rh/r},m_\chi,m_S,m_N), &n>5
\end{cases}\\
&T_{FI} \big|^{\text{EMDE}} \sim \begin{cases}
\text{max}(T_e,m_\chi,m_S,m_N), &n<4.5 \\
\text{max}(T_i,m_\chi,m_S,m_N), &n>4.5
\end{cases}\\
&T_{FI} \big|^{\text{EP}} \sim \begin{cases}
\text{max}(T_r,m_\chi,m_S,m_N), &n<12 \\
\text{max}(T_e,m_\chi,m_S,m_N), &n>12\,.
\end{cases}
\end{aligned}
\end{equation}

In our analysis, we neglect the subdominant contribution of production during the post-inflationary reheating period (above $T_\rh$), as it would be only important for models featuring $R_\chi \propto T^n$ with $n>12$ or with mediator masses above $T_\rh$.

\begin{figure}[!t]
\centering
\includegraphics[scale=0.61]{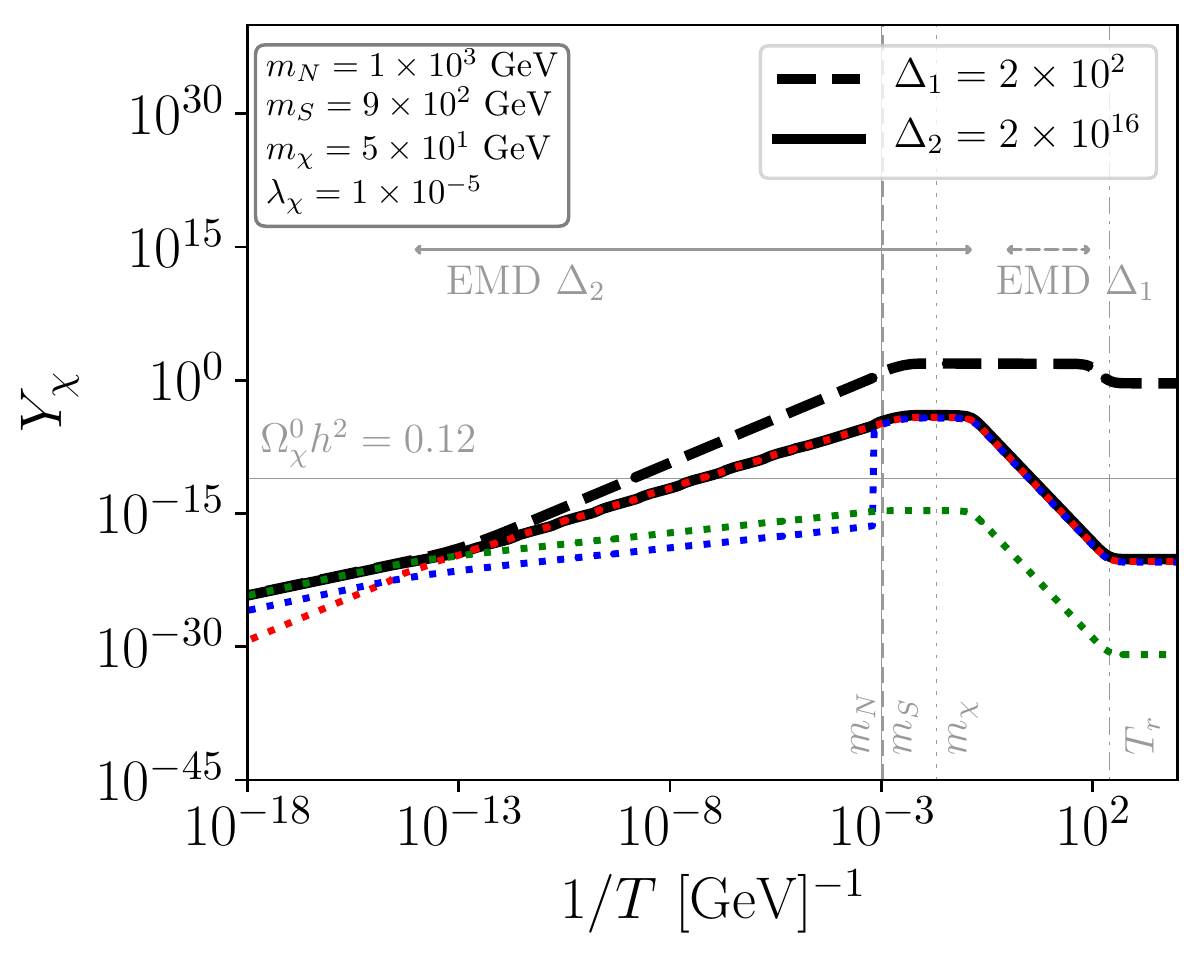}
\includegraphics[scale=0.61]{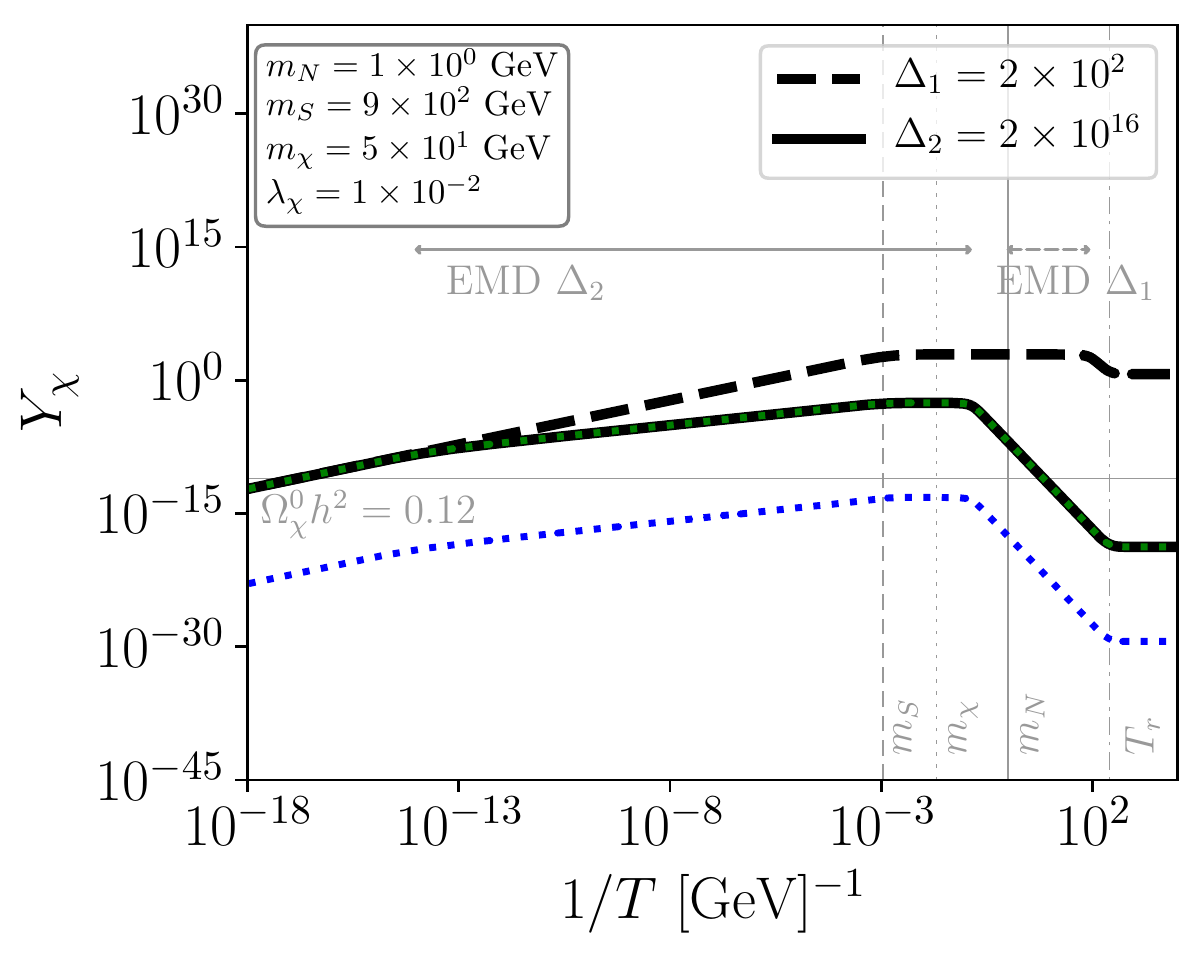}
\caption{
Evolution of the yield of $\chi$ in the case of small and large entropy production (dashed and solid curves respectively). In the left (right) panel we have $m_N > m_\chi, m_S$ ($m_N < m_\chi, m_S$). For larger entropy production ($\Delta = 2\times10^{16}$), contributions from decay (dotted red), s-channels (dotted blue) and t-channels (dotted green) are also shown. 
}
\label{Fig:evolution}
\end{figure}

In~\autoref{Fig:evolution} we show the solution of the coupled system of Boltzmann fluid equations for the evolution of $\chi$ (\autoref{equ:DMevol}), the generic matter content $M$ and radiation $R$, as discussed in~\autoref{sec:early} and detailed in~\autoref{sec:freezein_evolution}. We choose $m_N > m_\chi$ in the left panel and $m_N < m_\chi$ in right panel, with detailed values shown in the plots. We report our results for two possibilities regarding the dilution factor $\Delta$. The case of a short EMDE era with $\Delta = \Delta_1 = 2\times 10^2$ is indicated by the dashed grey arrows (for temperatures from $T_i = 1$\,GeV to $T_e = 1.2 \times 10^{-2}$\,GeV) and corresponds to the dashed black curves, whereas the case of a long EMDE is indicated by the solid grey arrows (for temperatures from $T_i = 10^{14}$\,GeV to $T_e = 7.6$\,GeV), with $\Delta = \Delta_2 = 2\times 10^{16}$, corresponding to the solid black curves. In the latter case, we show explicitly the contributions of the decay, s-channel and t-channel processes for the relic density of $\chi$ (red, blue and green dotted curves respectively). In both cases, the reheat temperature of the EMDE is set as $T_r = 4 \times 10^{-3}$\,GeV, so that the dilution of the relic density can be observed from the corresponding values of $T_e$ until $T_r$.

The decay process becomes inefficient at temperatures just below the decaying field mass, so that its contribution for the relic density levels off around $T\sim m_N$. As we can see, the on-shell production of $N$ from $l H$ annihilations around $T\sim m_N$ lead essentially to the decay contribution. Finally, we can also see that the t-channel becomes ineffective around the Boltzmann suppression of $N$.

\subsection{Thermalization of \texorpdfstring{$N$}{N} during the DM Freeze-in Production}
\label{sec:thermalization_N}

Since the freeze-in happens when thermal bath species produce dark matter in out-of-equilibrium processes, it is important to know under which conditions the heavy neutrinos thermalize with the SM particles during the DM freeze-in production.

The chemical equilibrium of heavy neutrinos with the SM bath would have been mainly driven by decays and inverse decays involving Higgs and leptons ($N \tob H l$ or $H \tob N l$)~\cite{Yaguna:2008mi,Hambye:2016sby}, whichever kinematically available, since they are proportional to $|Y^{ij}_\nu|^2$. In our analysis we are not going to consider the subdominant t-channel annihilations $NN\to ll$ and $NN\to HH$, both proportional to $|Y_\nu^{ij}|^4$.

To determine the thermalization condition, we compare the heavy neutrino or the Higgs decay width with the Hubble parameter in the corresponding phase of the Universe's evolution (\autoref{equ:Hubble_Final_ERD}-\autoref{equ:Hubble_Final_RD}). Thus, when $\Gamma > H(T)$ at temperatures relevant to the freeze-in production, the heavy neutrinos are thermalized. As a consequence of the heavy neutrinos being thermalized, all the processes considered in the previous section contribute for the production of $\chi$ or $S$. Otherwise, the heavy neutrinos are not abundant enough as to  effectively decay and annihilate via t-channel into FIMPs and only the s-channel annihilation of Higgs or gauge bosons and leptons can be considered\footnote{In this case, $N$ will be produced via the freeze-in mechanism and then it will contribute to the DM relic abundance through its decay. We temporarily ignore this contribution. Solving the coupled Boltzmann equations is needed for a concrete treatment, which is beyond the scope of this work.}.

\begin{figure}[!t]
\centering
\includegraphics[width=0.49\textwidth]{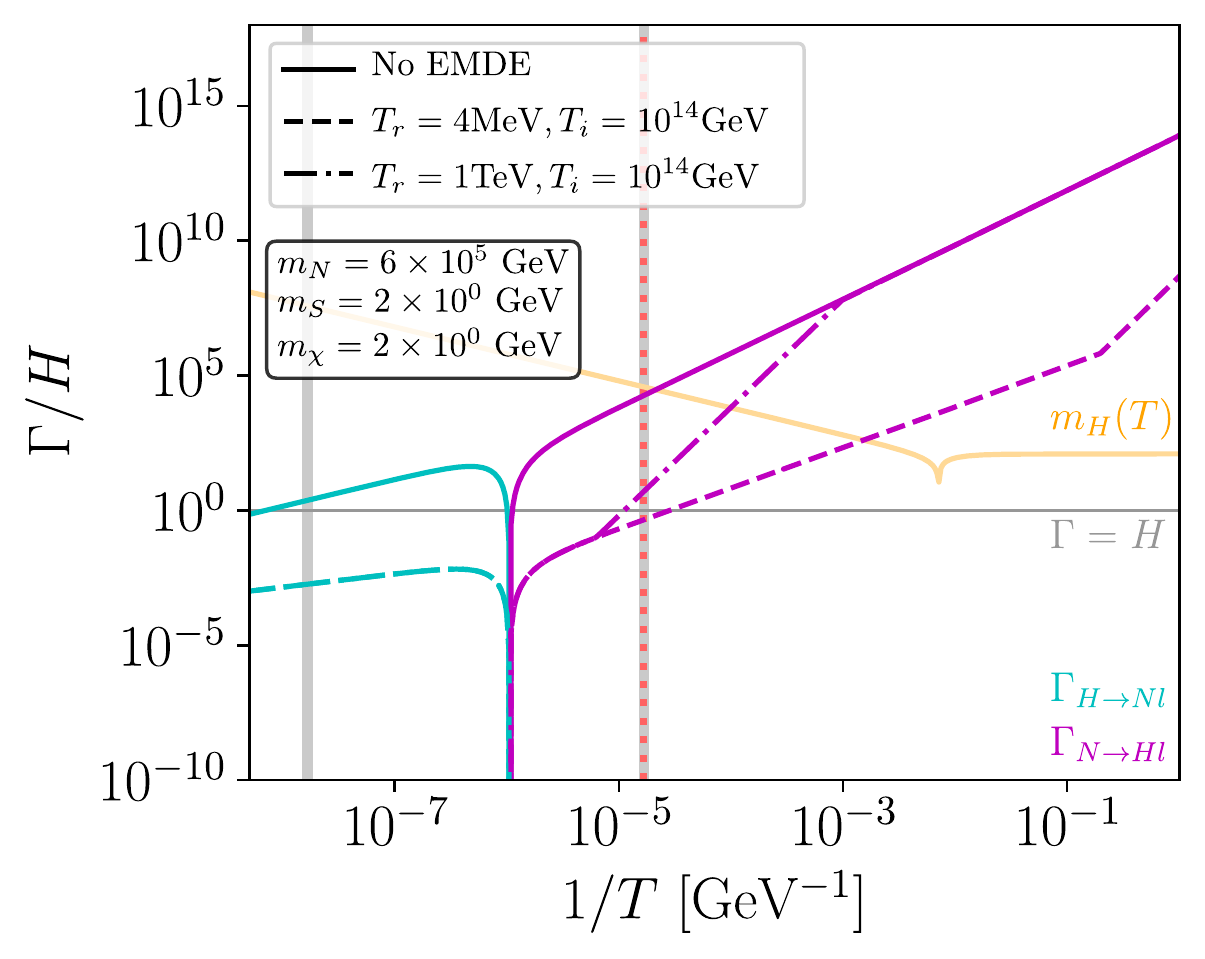}
\put(-25,150){(a)}
\includegraphics[width=0.49\textwidth]{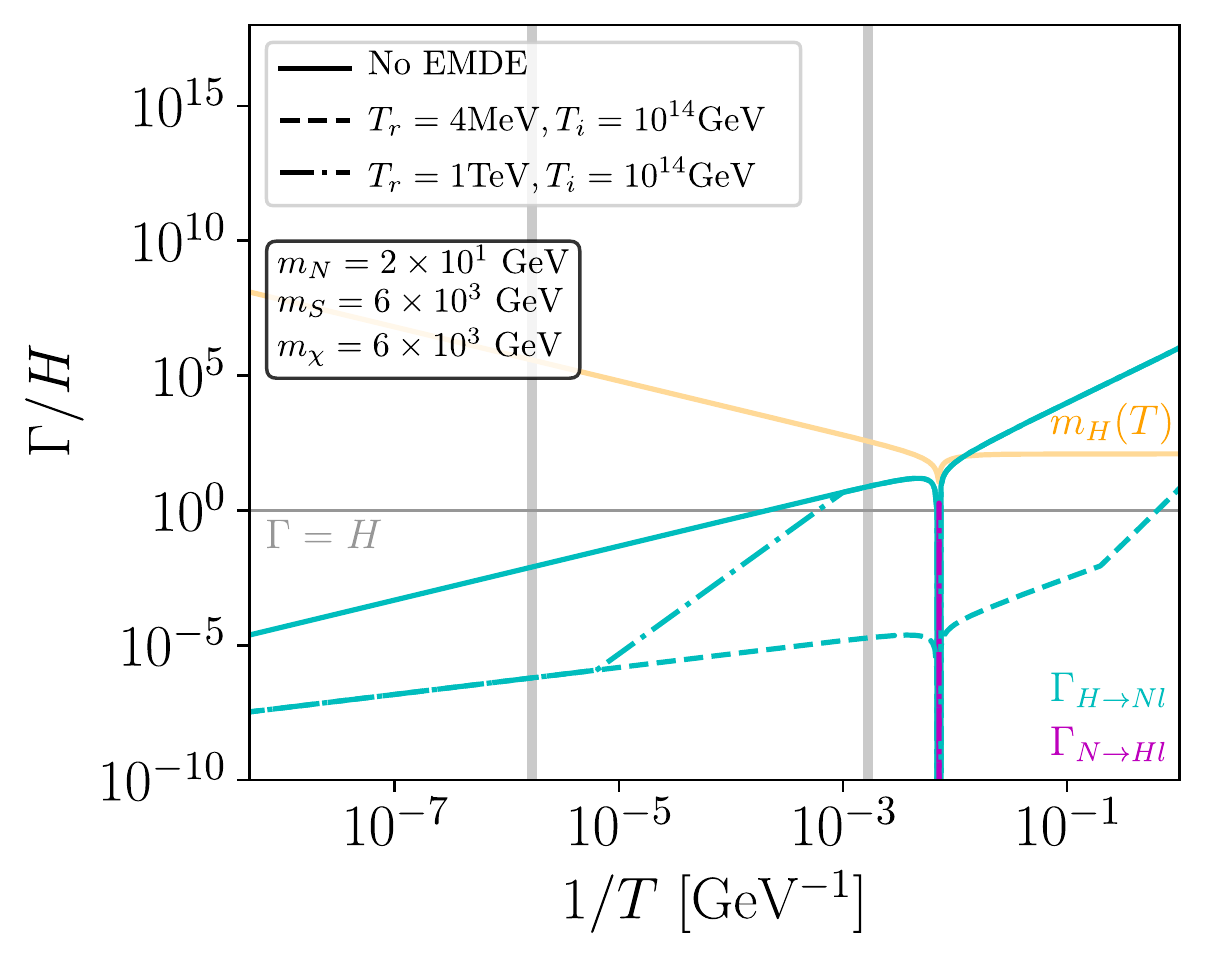}
\put(-25,150){(b)}\\
\includegraphics[width=0.49\textwidth]{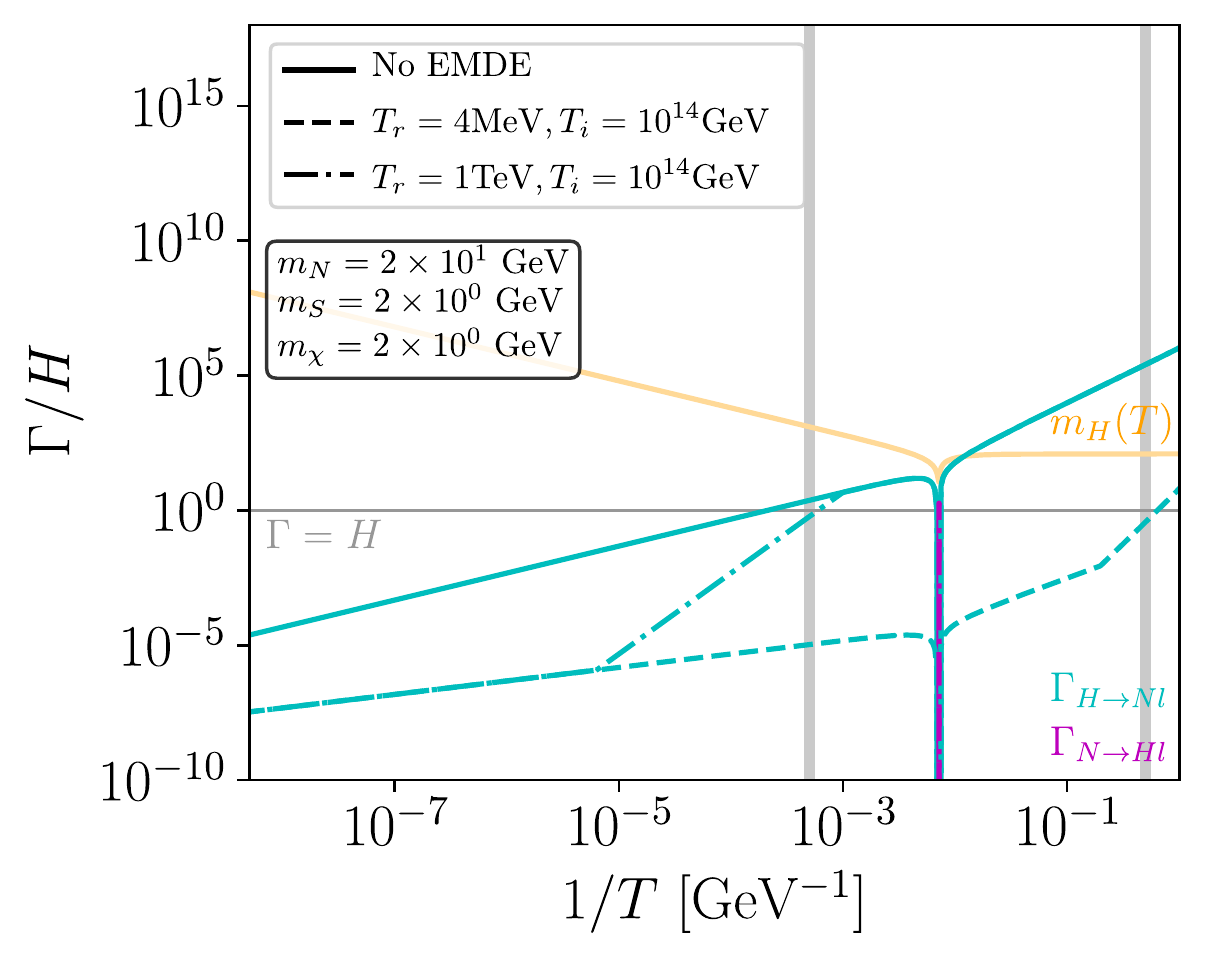}
\put(-25,150){(c)}
\caption{$\Gamma/H$ for the Higgs (cyan) and heavy neutrino (magenta) decays, in three cases regarding the EMDE. The period between gray vertical lines is relevant for the freeze-in processes. For temperatures below the red dotted line, $N$ leaves the thermal bath. For reference, we display the temperature-dependent Higgs mass (yellow curve). \textit{Panel (a):} $m_N \gg m_S, m_\chi$, $N$ not thermalized only for long enough EMDE. \textit{Panel (b):} $m_N \ll m_S, m_\chi$, $N$ never thermalized at relevant temperatures. \textit{Panel (c):} $m_N > m_S, m_\chi$, $N$ thermalized for short enough EMDE or for high enough $T_r$.}
\label{Fig:thermalization}
\end{figure}

In~\autoref{Fig:thermalization}, we show representative cases of the ratios of decay widths $\Gamma_{N\to Hl}(T)$ (magenta curves) and $\Gamma_{H\to Nl}(T)$ (cyan curves) to the Hubble rate $H(T)$, as well as the temperature-dependent Higgs mass $m_H(T)$ (yellow curves) as function of the inverse of temperature. Three cosmological scenarios are considered: 
\begin{itemize}
\item without an EMDE (solid curves);
\item EMDE with $T_r = 4$\,MeV, $T_e = 5$\,GeV and $T_i = 10^{14}$\,GeV (dashed curves);
\item EMDE with $T_r = 1$\,TeV, $T_e = 1.6 \times 10^5$\,GeV and $T_i = 10^{14}$\,GeV (dot-dashed curves).
\end{itemize}

We can recognize in~\autoref{Fig:thermalization} different slopes in the $\Gamma/H$ ratios, which are mainly due to the different temperature dependencies of the Hubble rate. Along the solid lines, the Universe is always RD. The abrupt changes in the slope observed in the dashed and dot-dashed curves happen once the Universe goes from an EMDE to an EP period, towards the slope expected for a RD era, below the respective $T_r$ values. On the other hand, the height of each curve for the same cosmic period depends on the Yukawa coupling $Y_\nu^{ij}$, which is proportional to $\sqrt{m_N}$.

We check whether $\Gamma/H>1$ over temperatures within the vertical gray lines, from max$(m_N, m_S, m_\chi)/10$ to $100\times$max$(m_N, m_S, m_\chi)$, which is an interval relevant for the DM freeze-in production. Roughly below $T\sim m_N$, the abundance of $N$ becomes Boltzmann-suppressed. Hence, we set a conservative lower bound on $T$ at $m_N/10$, as indicated by the vertical dotted red line, if the decay of $N$ (and back reaction) is relevant for its thermalization. Once the abundance of the Higgs boson becomes Boltzmann-suppressed, though, scattering with leptons could keep $N$ in the thermal bath. Hence, whether or not $N$ is thermalized during DM freeze-in production highly depends on the relation among $m_N$ (which sets the Yukawa coupling), max$(m_N, m_S, m_\chi)$ (which determines the freeze-in production temperature) and $T_i/T_r$ (which determines the cosmic history).

Panels (a) and (b) of~\autoref{Fig:thermalization} display two limiting cases regarding the mass of the heavy neutrinos: $m_N \gg m_S, m_\chi$ and $m_N \ll m_S, m_\chi$, respectively. When $m_N\gg m_S, m_\chi$ (panel (a)), the DM freeze-in production  happens mainly at temperatures close to $m_N$. A relatively large mass of $N$ will result in a large Yukawa coupling which, in turn, yields a corresponding large decay width. Hence, it is easier for $N$ to thermalize with the cosmic bath and a long enough EMDE is required to avoid the thermalization (dashed and dot-dashed curves). On the other hand, when $m_N \ll m_S, m_\chi$, the DM freeze-in production happens at temperatures much higher than $m_N$. In this case, the decay width of the heavy neutrino/Higgs is suppressed by the Yukawa coupling,  which hampers the thermalization of $N$, regardless the duration of the EMDE (panel (b)). However, for light $N$,  relatively small $\chi$ and $S$ masses can help achieving the heavy neutrino thermalization during the DM freeze-in production,  by lowering the freeze-in temperature (solid curve), as shown in the panel (c) of~\autoref{Fig:thermalization}. In general, a long EMDE hampers the thermalization process, since the Universe expands faster during this period, as we have pointed out before. Nevertheless, choosing a different $T_r$ (recall that this is the temperature at the end of the EMDE), the thermalization of $N$ can still be achieved over DM freeze-in production temperatures, as shown by the dot-dashed curve in the panel (c) of~\autoref{Fig:thermalization}.

In summary, for heavy $N$, the thermalization is easily attained and, therefore, a long EMDE is required to avoid it. In the scenario where $N$ is light, there are two possibilities: if $\chi$ and $S$ are heavy, it is hard to achieve thermalization, no matter how long the EMDE lasts. On the other hand, if $\chi$ and $S$ are also light, the thermalization is reached even for a long EMDE, as long as it ends around the temperature at which the freeze-in production becomes efficient.

\subsection{Freeze-in Conditions}
\label{sec:freezein_decoupling}

Let us now consider in more detail the conditions for the freeze-in regime to hold. Besides the assumption of negligible initial abundance, the production of bath species from FIMPs needs to be avoided. To ensure that FIMPs are never as abundant as bath species, our parameter space is such that the reaction rates of FIMP production from the thermal bath fields $i$ ($R(T)/n_i(T)$) were always slower than the cosmic expansion rate ($H(T)$).

If the heavy neutrinos were always coupled to the thermal bath at the relevant temperatures for the freeze-in processes, all the following conditions need to be satisfied:
\begin{equation}\begin{split}
[I]& \hspace{1cm} \frac{\Gamma_{N \to \bar\chi S}}{H(T)} \ll 1,
\\
[II]& \hspace{1cm} \frac{n_{H^+}\la \sigma v \ra_{l_i H^+ \to \bar\chi S}}{H(T)} \ll 1  ~\text{and}~ \frac{n_{H^0}\la \sigma v \ra_{\nu_i H^0 \to \bar \chi S}}{H(T)} \ll 1, \\
[III]& \hspace{1cm} \frac{n_{N}\la \sigma v \ra_{N N \to \bar\chi \chi}}{H(T)} \ll 1 ~\text{or}~ \frac{n_{N}\la \sigma v \ra_{N N \to S^* S}}{H(T)} \ll 1.
\end{split}
\label{FI cond}
\end{equation}

Notice that we will have different conditions for the different eras in the cosmic history, so that we need to consider the corresponding expression for $H(T)$ at each period (\autoref{equ:Hubble_Final_ERD}- \autoref{equ:Hubble_Final_RD}).

\begin{figure}[!t]
\centering
\includegraphics[scale=0.59]{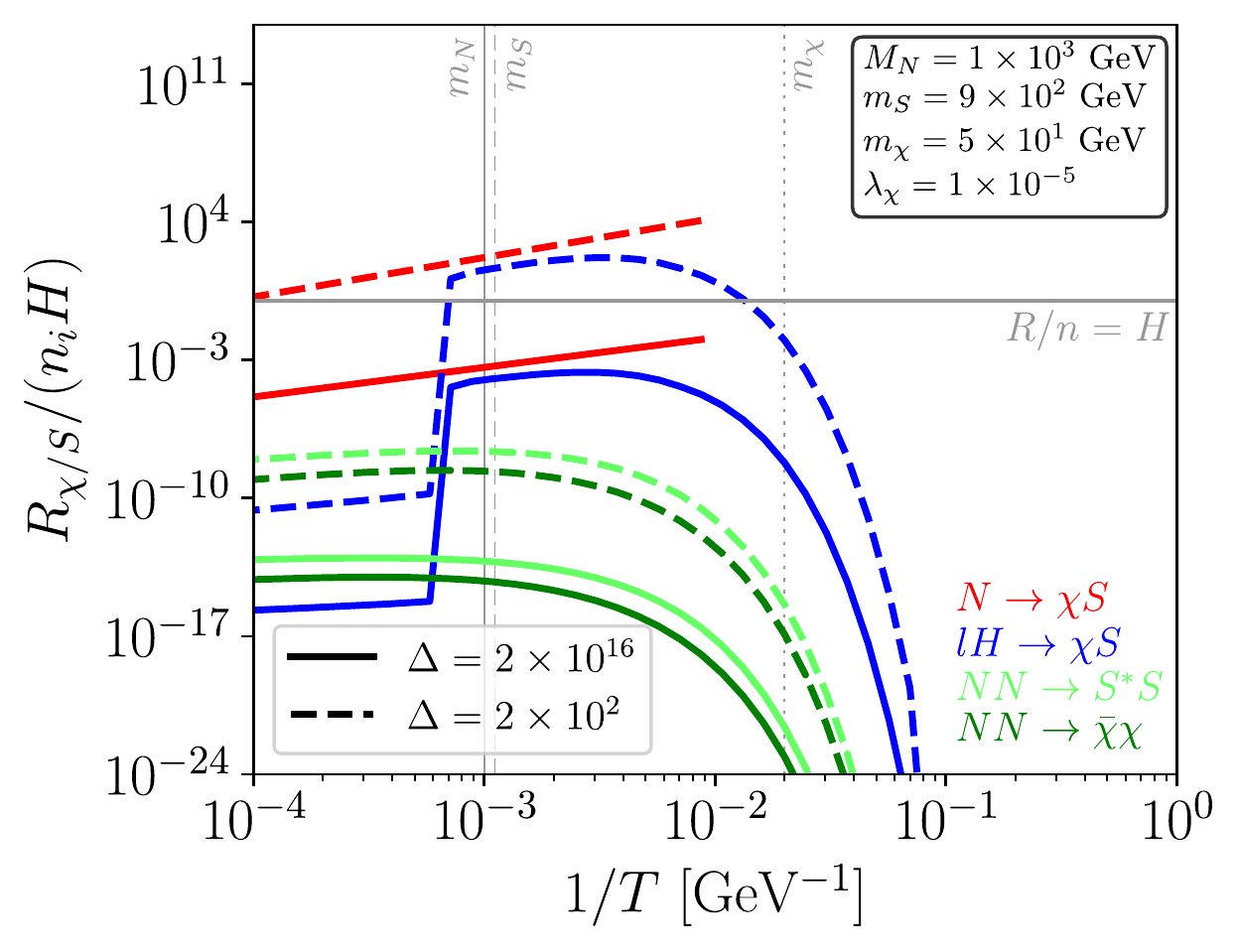}
\includegraphics[scale=0.59]{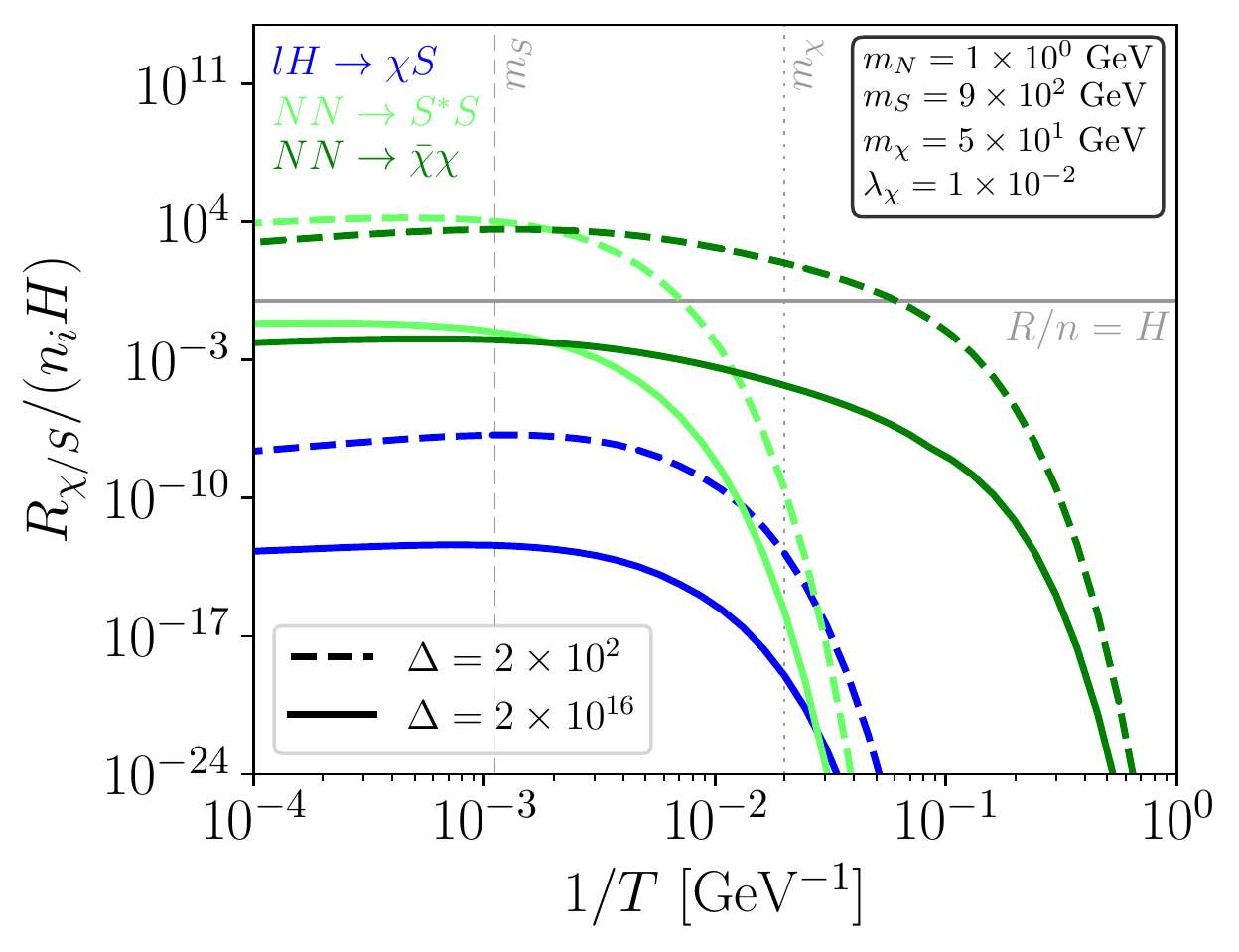}
\caption{Ratios between the DM production and the Hubble rates for the decay (red), s-channels (blue) and t-channels (green), in the case of small and large entropy production (dashed and solid lines, respectively). In the left (right) panel we have $m_N > m_\chi, m_S$ ($m_N < m_\chi, m_S$). Notice that, the longer the EMDE, the easier to satisfy the out-of-equilibrium conditions, allowing for larger couplings in the freeze-in regime. 
}
\label{Fig:decoupling}
\end{figure}

In~\autoref{Fig:decoupling}, we show the ratios between the production rates and the appropriate Hubble expansion rate for the scenario in which $m_N = 1000 \,\text{GeV} > m_\chi, m_S$ (left panel) and $m_N = 1 \,\text{GeV} < m_\chi, m_S$ (right panel), with $m_S = 900$\,GeV and $m_\chi = 50$\,GeV. In this plot, we have set the reheating temperature to be $T_r = 4$\,MeV and considered two possible values of $T_i$: $T_i = 1$\,GeV (dashed curves), corresponding to an entropy production of $\Delta \simeq 2\times 10^2$, and $T_i = 10^{14}$\,GeV (solid curves), corresponding to an entropy production of $\Delta \simeq 2\times 10^{16}$. Therefore, in the dashed curves the Hubble rate is dominated by radiation, while in the solid curves it is dominated by matter.

In the case of the decay (red curves), the production rate is given only by $\Gamma_{N \to \chi S}$, which does not depend on the temperature. The slightly different slopes we see are only due to the different Hubble rates. The blue curves correspond to the s-channel contributions, so that $R_{\chi/S}/(n_H H) = n_H \la \sigma v \ra / H$. Around $T \sim m_N$, as we have said, $N$ is produced on-shell and $n_H \la \sigma v \ra / H \sim \Gamma_{N \to \chi S}/H$ and, for lower temperatures, $n_H \la \sigma v \ra / H$ becomes ineffective due to the Boltzmann suppression on $n_H$. The t-channel contributions shown in green, with $R_{\chi/S}/(n_N H) = n_N \la \sigma v \ra / H$, become ineffective due to the Boltzmann suppression of the initial or final states. While in this figure we have integrated~\autoref{rate density scatt} without any approximation, \autoref{equ:NWA}, \autoref{rateschannel}
and~\autoref{ratetchannel} were found to be fair approximations.

In the case where the heavy neutrinos were never coupled to the thermal bath, they were always much less abundant than SM species. Therefore, their decays and t-channel annihilations were always negligible with respect to the SM s-channel annihilations and only the condition [II] of~\autoref{FI cond} is to be ensured.

In our numerical scans of~\autoref{sec:results} and~\autoref{sec:pheno}, the conditions~\autoref{FI cond} are always satisfied. As an example of how they constrain our parameter space, let us consider the decay channel [I]. At temperatures $T \lesssim m_N/10$, the decay channel becomes negligible due to the Boltzmann suppression on the abundance of $N$. For this reason, the out-of-equilibrium condition for the decay $N \to \bar\chi S$ can be set at the minimum temperature $T \approx m_N/10$. Considering the parameters of~\autoref{Fig:decoupling}, the out-of-equilibrium condition [I] reads:
\begin{equation}\begin{split}\label{FIconditions}
\lam \ll &\left(\frac{10^3\,\gev}{m_N}\right)^{\frac{1}{2}} \left(\frac{g_e(100 \,\gev)}{103.5}\right)^{\frac{1}{4}} \frac{0.01}{(1-\epsilon^2)} \\
&\times 
\begin{cases}
2.5 \times 10^{-8}\,  \frac{T}{100\,\gev}\,, ~~&\text{for}~ \Delta = 1\\
1.5 \times 10^{-4}\, \left(\frac{T}{100\,\gev}\right)^{\frac{3}{4}} \left(\frac{T_r}{4\,\text{MeV}}\right)^{\frac{1}{4}} \left(\frac{\Delta}{2\times 10^{16}}\right)^{\frac{1}{4}}\,,  ~~&\text{for}~ \Delta = 2 \times 10^{16}
\end{cases}
\end{split}\end{equation}

Thermal equilibrium between $\chi$ and $S$ could be achieved via $\chi S \to S \chi$ and $\chi \chi \to S S$, with t-channel exchanges of heavy neutrinos. Since these processes are proportional to $\lambda_\chi^4$, which we are already constraining with the freeze-in conditions, we can safely assume that our FIMP candidates $\chi$ and $S$ do not constitute a decoupled thermal bath.

We can therefore conclude that the possibility of a long EMDE \textit{allows} for out-of-equilibrium processes with larger couplings. Interestingly, as we are going to see in the next section, larger couplings are \textit{needed} for keeping the correct value for the FIMP relic density in the case of longer EMDE. Having seen how the early matter era affects the relic density of our dark matter candidate, in the next section we show the parameter space of our model providing the right amount of relic density either for $\chi$ or $S$.

\section{Study of the Parameter Space}
\label{sec:results}

In this section, we study the regions of our free parameter space -- comprised of $m_\chi$, $m_N$, $m_S$, $\lambda_\chi$, $T_r$ and $T_i$ -- providing the correct present relic density of $\Omega^0 h^2 \simeq 0.12$~\cite{Aghanim:2018eyx} for the fermionic dark matter candidate $\chi$. Results for the scalar dark matter do not significantly differ from those shown here. 

As we have discussed in~\autoref{sec:freezein}, non-thermalized heavy neutrinos are always much less abundant than the SM species and are not able to efficiently produce dark matter. In this work, we roughly take this into account by considering the contributions of decays and t-channels for the evolution of the DM relic density only when the heavy neutrinos thermalize reasonably before the freeze-in time. Otherwise, we consider only the s-channel contributions since SM species are the dominant ones. Carefully taking this issue into account, by tracking the coupled evolution of $\chi$ and $N$, is beyond the scope of this work.

\begin{figure}[!t]
    \centering
\includegraphics[width=\textwidth,trim=150 0 150 0]{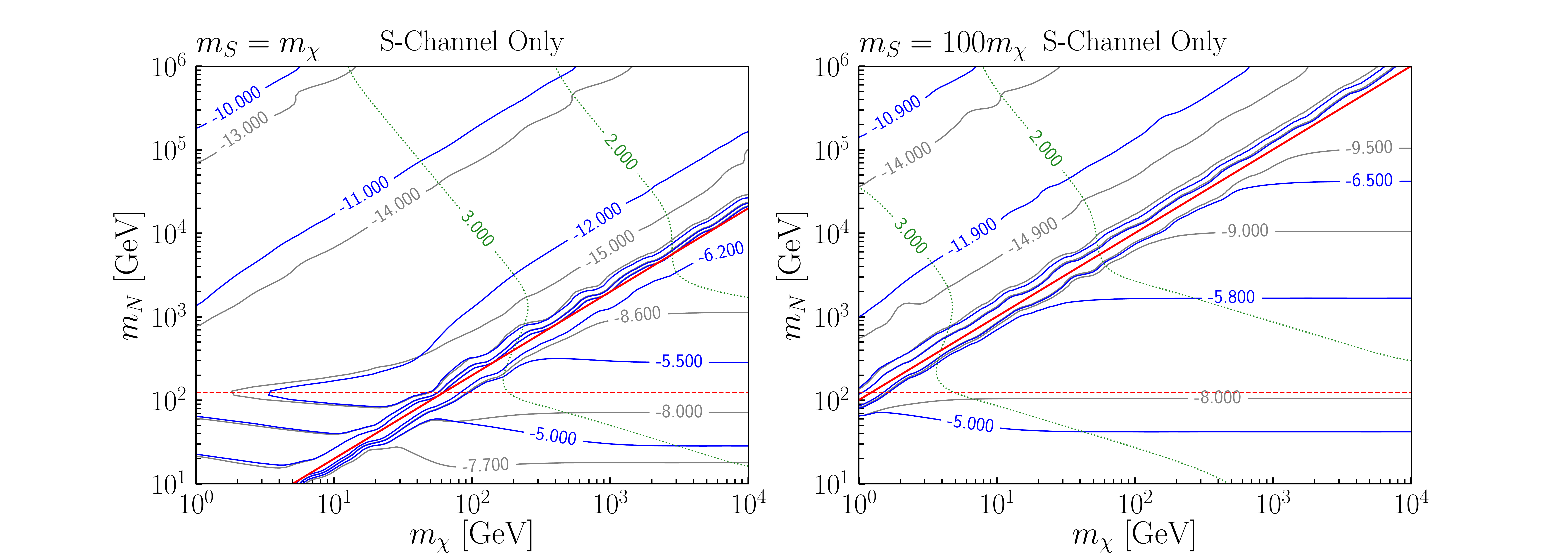}
\caption{Contours of $\lambda_\chi$ (in logarithmic scale) providing the correct relic density for $\chi$ in the $m_\chi$-$m_N$ plane, when only s-channels are considered. The gray, blue and green contours are for $T_i = 10$\,MeV, $10^3$\,GeV and $10^{14}$\,GeV, respectively, with $T_r = 10$\,MeV. Solid (dashed) red lines are for $m_N = m_S + m_\chi$ ($m_N = m_h = 125$ GeV).}
\label{fig:mchi-mN-lam-contour_schan}
\end{figure}

\begin{figure}[!t]
    \centering
    \includegraphics[width=\textwidth,trim=150 0 150 0]{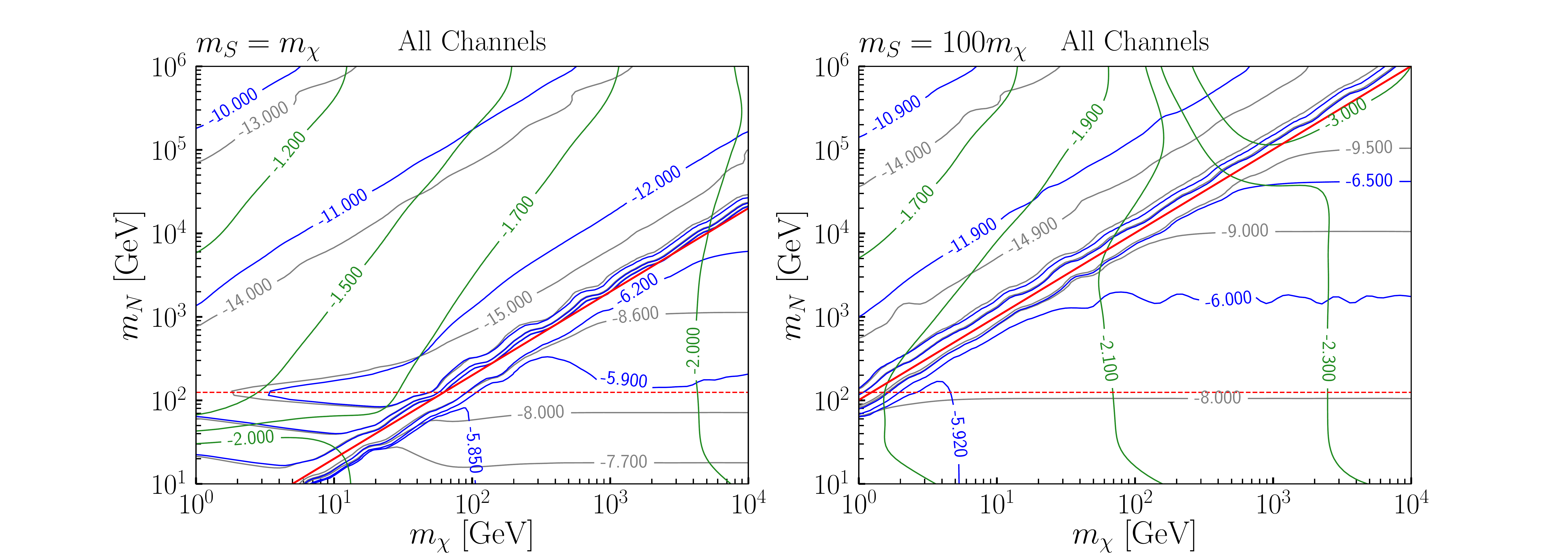}
    \caption{Same as~\autoref{fig:mchi-mN-lam-contour_schan}, but considering decays, s-channels and t-channels for the establishment of the relic density of $\chi$. Solid (dashed) red lines are for $m_N = m_S + m_\chi$ ($m_N = m_h = 125$ GeV).}
    \label{fig:mchi-mN-lam-contour_all}
\end{figure}

In order to understand the impact of this assumption, we show in~\autoref{fig:mchi-mN-lam-contour_schan} and~\autoref{fig:mchi-mN-lam-contour_all} the contours of $\lambda_\chi$ providing the correct relic density of $\chi$ in the plane $m_\chi$-$m_N$, considering respectively only s-channels and all the channels. In these plots, we \textit{do not} check the thermalization of heavy neutrinos nor the freeze-in conditions. We investigate the effect of the duration of the EMDE by considering three different cases: 
\begin{itemize}
\item without an EMDE (gray curves);
\item EMDE from $T_i = 10^3$\,GeV to $T_e \simeq 110$\,MeV, with $T_r = 10$\,MeV (blue curves);
\item EMDE from $T_i = 10^{14}$\,GeV to $T_e \simeq 24$\,GeV, with $T_r = 10$\,MeV (green curves).
\end{itemize}

The hierarchy between $m_N, m_S$ and $T_i$ is studied by setting $m_S = m_\chi$ in the left panels and $m_S = 100 m_\chi$ in the right panels. The region for $m_\chi > m_N$ is of particular interest since it is when the DM t-channel annihilations relevant for indirect detection searches become possible, as we explore in the next section.

Let us first focus on~\autoref{fig:mchi-mN-lam-contour_schan}, where we only consider the contribution from s-channels in the freeze-in process. In order to understand the features in~\autoref{fig:mchi-mN-lam-contour_schan}, the approximations for the s-channel reaction rate density in different regions can be helpful. In the resonant region, where the mediator $N$ can be produced on-shell ($m_N > m_S + m_\chi$ and $m_N > m_H(T) + m_\ell \gtrsim m_h = 125$ GeV), from~\autoref{equ:NWA}, we have: 
\begin{align}
    R_\chi(T) \Big|^{\rm on-shell} &\propto |\lam|^2 m_N^4 T \frac{|Y_\nu^{ij}|^2}{\Gamma_N}, 
\end{align}
where we have neglected some irrelevant factors related to the phase space and distributions, and  $\Gamma_N$ is the total width of $N$, which depends on both $|\lam|^2$ and $|Y_\nu^{ij}|^2$ (see~\autoref{sec:amplitudes} for a detailed expression of $\Gamma_N$). On the other hand, in the non-resonant region, where the mediator $N$ can only be produced off-shell ($m_N < m_S + m_\chi$ or $m_N < m_h = 125$ GeV), from~\autoref{rateschannel}, we have:
\begin{align}
    R_\chi(T)\Big|^{\rm off-shell} &\propto |\lam|^2|Y_\nu^{ij}|^2T^4.
\end{align}

In the case of a RD Universe, along the gray and the blue lines\footnote{Along the blue lines, there will be a period of EMDE, however, with $T_i = 1$ TeV, the effects will be small.}, the DM yield $Y^0_\chi$ in~\autoref{equ:Y0} can be estimated by only $y_{\RD}$, integrating from $m_N$ (resonant region) or $m_\chi$ (non-resonant region) to $T_{\rh}$. Therefore, it follows, approximately:
\begin{align}
    \label{equ:approx_RD}
    \Omega h^2 &\propto m_\chi \int dT\frac{R_\chi(T)}{T^6}\nonumber \\
    &\sim \begin{cases}
     |\lam|^2m_\chi m_N^4\frac{|Y_\nu^{ij}|^2}{\Gamma_N}\int_{m_N}^{T_\rh}\frac{dT}{T^5}\approx |\lam|^2\frac{m_\chi}{m_N}   & \text{resonant region}\\
     |\lam|^2m_\chi m_N \int_{m_\chi}^{T_\rh}\frac{dT}{T^2} \approx |\lam|^2m_N & \text{non-resonant region,}
    \end{cases}
\end{align}
where we have used $|Y_\nu^{ij}|\sim \sqrt{m_\nu m_N}/v$ and, in the resonant region, for small enough $|\lam|$, $\Gamma_N\propto m_N|Y_\nu^{ij}|^2$. From the two approximations in~\autoref{equ:approx_RD}, we can see that, for grey and blue lines, in the resonant region, the slope for each equal-$\lam$ contour will be 1,  while in the non-resonant region, it will be 0 (almost no dependence on $m_\chi$). It is also interesting to notice that, in the resonant region, a smaller coupling is required to achieve the correct relic abundance, when compared to the non-resonant region, due to the enhancement of the cross-section.

For much larger $T_i$, as along the green lines, the contribution to the DM yield from the EMDE is dominant. Thus, we have:
\begin{align}
    \label{equ:approx_EMDE}
    &\Omega h^2 \propto m_\chi \int dT\frac{R_\chi(T)}{T^{11/2}}\nonumber \\
    &\sim \begin{cases}
     |\lam|^2m_\chi m_N^4\frac{|Y_\nu^{ij}|^2}{\Gamma_N}\int_{m_N}^{T_\rh}\frac{dT}{T^{9/2}}\approx \frac{|\lam|^2}{\Gamma_N/m_N}m_\chi\sqrt{m_N}\sim f(\lam)m_\chi\sqrt{m_N}   & \text{resonant region}\\
     |\lam|^2m_\chi m_N \int_{m_\chi}^{T_\rh}\frac{dT}{T^{3/2}} \approx |\lam|^2m_N\sqrt{m_\chi} & \text{non-resonant region,}
    \end{cases}
\end{align}
where, for the resonant region, we have used an unknown function to represent the dependence on $\lam$, since, for large $\lam$,  the NWA is no longer a good approximation. Besides this caveat, from the two approximations in~\autoref{equ:approx_EMDE}, we can see that, for the green lines, in the resonant region, the slope for each equal-$\lam$ contour will be -2, whereas in the non-resonant region it will be -1/2. Also note that, along the green lines, the continuity of the couplings between resonant region and non-resonant region is also an indication of the failure of NWA.

Let us now turn our attention to the case where also decays and t-channels are considered (\autoref{fig:mchi-mN-lam-contour_all}). While analytic approximations would be much more complicated as the t-channel contribution is more complex, we can take important information from this figure. We can clearly see that increasing values of $T_i$ make the t-channel to play an increasing role in the freeze-in, as $\lam$ needs to be larger. When $m_N$ is heavy enough so that $Y_\nu^{ij}$ is sizable, the s-channels ($\propto |\lam|^2 |Y_\nu^{ij}|^2$) dominate over decays ($\propto |\lam|^2$) and t-channels ($\propto |\lam|^4$). We can observe this along the gray and on-shell region of blue contours. For $T_i = 1$\,TeV (blue), we see that the t-channel starts to dominate only in the ($Y_\nu^{ij}$-suppressed) non-resonant region (its slope is different from the s-channel only cases).
In the case of $T_i = 10^{14}$\,GeV (green), the need for much larger couplings makes the t-channel to always dominate in our parameter space. In this case, though, $\lam$ does not need to be extremely large as in the case where only s-channels contribute to the freeze-in.

\section{Phenomenology}
\label{sec:pheno}

Having discussed the dynamics of the model throughout the modified cosmic history and determined how the dark particle can account for the present dark matter abundance, in this section, we explore different methods to probe the model, which includes searching for signatures in direct and indirect detection experiments. For this purpose, we scan the parameter space for two cases: Case A, where $\chi$ is the dark matter; and Case B, where $S$ is the dark matter. The scanned ranges for all input parameters are listed in~\autoref{tab:inputranges}. Note that the three heavy neutrinos are chosen to be degenerate and the Yukawa interaction matrix, $Y_\nu^{ij}$, is fully determined by the heavy neutrino mass and the complex orthogonal matrix $R$, which is chosen to be the identity $\mathbb{I}$. Additionally, the freeze-in conditions are satisfied. The coupling $\lambda_\chi$ is chosen to provide the observed dark matter relic density for each parameter point. As pointed out in the  previous section, although a huge DM dilution requires couplings $\lambda_{\chi}>\mathcal{O}\left(1\right)$, in our plots, we consider only the parameter space where  $\lambda_\chi<4\pi$.

\begin{table}[!t]
    \centering
    %\resizebox{\textwidth}{!}{
    \begin{tabular}{|c|c|c|}
    \hline\hline
    Parameters & Case A & Case B \\
    \hline\hline
    $m_\chi$ &  [$1$ GeV, $10^4$ GeV] & [$m_S$, $10^6$ GeV] \\
    \hline
    $m_S$ &[$m_\chi$, $10^6$ GeV] & [$1$ GeV, $10^4$ GeV] \\
    \hline
    $m_N$ & \multicolumn{2}{c|}{[$10$ GeV, $10^6$ GeV]} \\
    \hline
    $T_i$ & \multicolumn{2}{c|}{[$10^2$ GeV, $5\times10^{14}$ GeV]} \\
    \hline
    $T_r$ & \multicolumn{2}{c|}{[4 MeV, $T_i$]} \\
    \hline\hline
    \end{tabular}%}
    \caption{The scan ranges for each input parameter in all cases. Note that $Y_\nu^{ij}$ is fully determined by $m_N$ and $R=\mathbb{I}$, and $\lambda_\chi$ is chosen to give the observed dark matter relic density and is required to be less than $4\pi$.}
    \label{tab:inputranges}
\end{table}

\subsection{Prospects for Direct Detection}
\label{sec:pheno_DD}

\begin{figure}[!btp]
    \centering
    \includegraphics[width=0.4\textwidth]{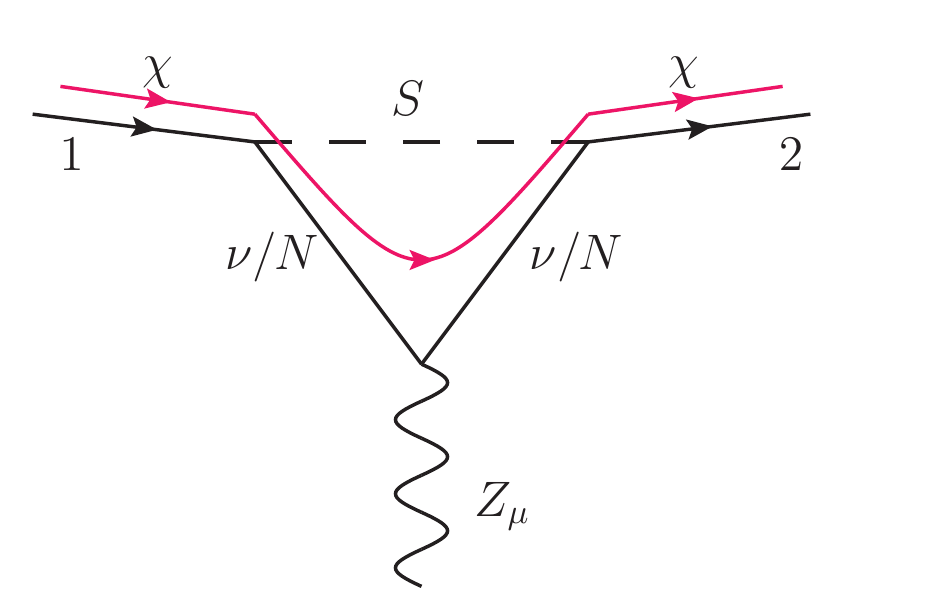}
    \includegraphics[width=0.4\textwidth]{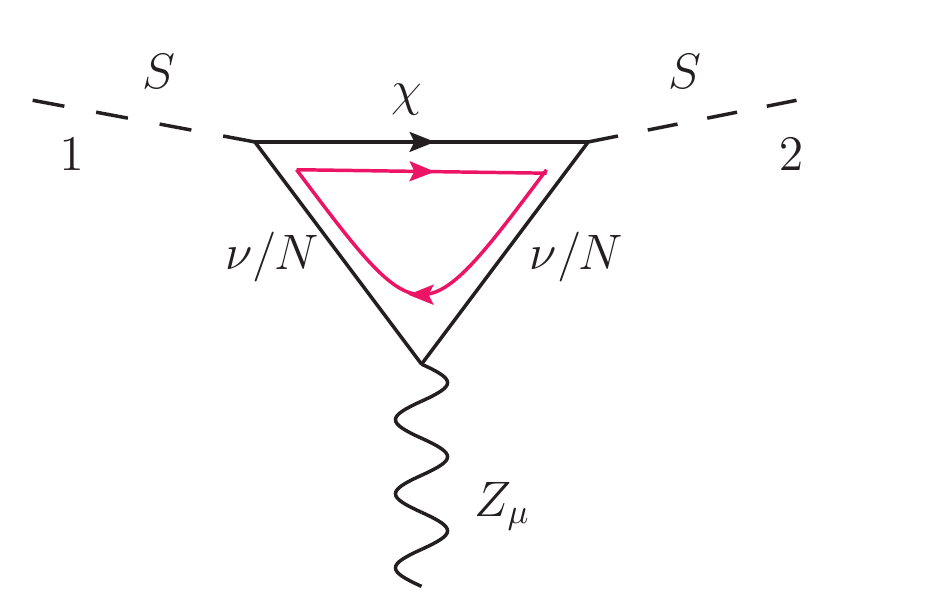}
    \caption{The loop induced vertices $Z-\chi-\chi$ (left) and $Z-S-S$ (right). For the process involving both Dirac and Majorana fermions, we use the convention in~\cite{Denner:1992vza} to calculate the amplitude. The red arrows represent the fermion flow along which the amplitude for the fermion line is calculated.
    }
    \label{fig:EffVertex}
\end{figure}

The direct detection experiments aim at identifying the deposited energies when the incident dark matter particle bombards the target nucleus of the detector and, then, is scattered off. In this model, the direct detection relevant vertices ($Z-\chi-\chi$ and $Z-S-S$) are induced from the loops shown in~\autoref{fig:EffVertex}. Since the vertex $Z-S-S$ is suppressed by momentum transfer when considered in the context of direct detection, we will not discuss the direct detection approach for Case B. For Case A, the effective $Z-\chi-\chi$ coupling is expressed as:
$\mathcal{L}\supset g_{Z\chi\chi} Z_\mu \overline{\chi}\gamma^\mu P_L\chi$, where:
\begin{align}
g_{Z\chi\chi} &= -\frac{g\,\lambda_\chi^2}{32\pi c_W}\sum_{ij}\left(V^\dagger V\right)_{ij}\left(2m_N^2C_0(m_\chi^2,0^-,m_\chi^2,m_S^2,m_N^2,m_N^2)\right.\nonumber\\
&\quad -m_N^2C_0(m_\chi^2,0^-,m_\chi^2,m_S^2,m_N^2,0) +4C_{00}(m_\chi^2,0^-,m_\chi^2,m_S^2,m_N^2,0)\nonumber\\
&\left.\quad -2C_{00}(m_\chi^2,0^-,m_\chi^2,m_S^2,m_N^2,m_N^2) -2C_{00}(m_\chi^2,0^-,m_\chi^2,m_S^2,0,0)\right),
\label{coupling_nuc}
\end{align}
where $g$ is the $SU(2)_L$ coupling, $c_W\equiv\cos\theta_W$ ($s_W\equiv\sin\theta_W$) is the cosine (sine) of the Weinberg angle, $C_0$ and $C_{00}$ are the loop functions in the convention of {\tt LoopTools}~\cite{Hahn:1998yk} and $0^-$ corresponds to the momentum transfer square in the scattering which is negative and close to zero. Note that the finiteness of the above expression is guaranteed by the unitarity of the mixing matrix $\mathbb{N}$.

Focusing on the spin-independent (SI) DM-nucleus scattering cross section, we have:
\begin{equation}
\begin{split}
\sigma_{0}^{SI} = & \frac{\mu_{\chi N}^2|g_{Z\chi\chi}|^2}{4\pi m_Z^4}\left[Z(g_d^V+2g_u^V) + (A-Z)(g_u^V+2g_d^V)\right]^2 \\
= & \frac{\mu_{\chi N}^2}{\mu_{\chi p}^2}{A^2}{\sigma_{\chi N}^{SI}},
\label{SI long}
\end{split}
\end{equation}
where $\sigma_{0}^{SI}$ is the SI DM-nucleus scattering cross section, $\sigma_{\chi N}^{SI}$ is the SI DM-nucleon scattering cross section, which is usually used to compare experimental results with different target isotopes, $\mu_{\chi N}$ is the reduced mass between DM and the nucleus (mass number A, charge number Z), $\mu_{\chi p}$ is the reduced mass between DM and the nucleon, i.e. $\mu_{\chi N}=\frac{m_\chi m_{\mathcal{N}}}{m_\chi+m_{\mathcal{N}}}$ and $\mu_{\chi p}=\frac{m_\chi m_p}{m_\chi+m_p}$ and $g_q^V$ are the corresponding vector current couplings of the quarks with Z-boson:
\begin{align}
g_u^V = \frac{g}{c_W}\left(\frac{1}{2} - \frac{4}{3}s_W^2\right), \quad
g_d^V = \frac{g}{c_W}\left(-\frac{1}{2} + \frac{2}{3}s_W^2\right).
\end{align}

\begin{figure}[!t]
    \centering
    \includegraphics[width=\textwidth]{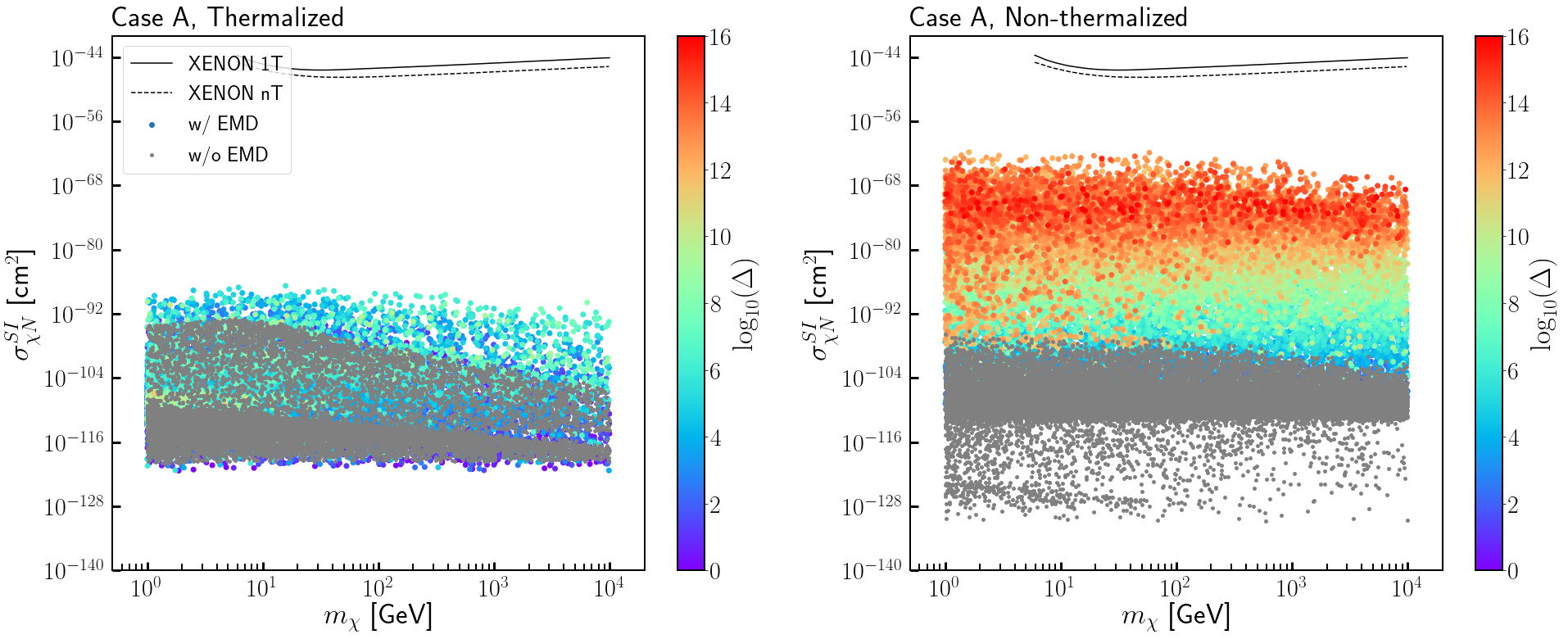}
    \caption{
        The SI scattering cross section with nucleon with (color points) and without (grey points) EMDE for Case A. The heavy neutrino is considered to be thermalized (left panel) and non-thermalized (right panel) respectively. The solid and dashed lines represent current {\tt XENON-1T} and prospected {\tt XENON-nT} bounds, respectively. 
        }
    \label{fig:directdetection}
\end{figure}

The SI cross section $\sigma_{\chi N}^{SI}$ is shown in~\autoref{fig:directdetection}, where each point provides the observed DM relic density in a scenario with (colored points) or without (grey points) EMDE. From~\autoref{fig:directdetection}, it is clear that the SI cross section is enhanced when the Universe undergoes an early matter era. It is also possible to see that the case where the heavy neutrinos are not thermalized with the cosmic plasma is more promising in terms of the possibility of being detected by direct searches in the future. This is due to the fact that, when the heavy neutrinos are not part of the thermal bath, the s-channel exchange of $N$ is the only mode that produces DM efficiently. Given that this channel is suppressed by the Yukawa couplings, the dark matter coupling $\lambda_\chi$ must be larger to achieve the observed DM relic abundance without compromising the freeze-in conditions. However, both cases presented in~\autoref{fig:directdetection} cannot be constrained by the most stringent bounds to date from {\tt XENON-1T}~\cite{Aprile:2017iyp} which, together with the projections from {\tt XENON-nT}~\cite{Aprile:2015uzo}, are also shown in~\autoref{fig:directdetection} for comparison. Even considering the ultimate liquid xenon direct detection experiment {\tt DARWIN}~\cite{Aalbers:2016jon}, or the liquid argon direct detection experiment {\tt ARGO}~\cite{Aalseth:2017fik}, which are the most promising experiments with a sensitivity for SI cross-sections $\sigma_{\chi N}^{SI}$ down to $\sim10^{-49}$ cm$^2$ (being able to reach the neutrino floor and to detect or exclude WIMPs with masses above 5 GeV), it would be hard to probe this model in the near future. However, note that, in our analysis, we use $R=\mathbb{I}$ as a conservative choice to focus on the effect only coming from cosmology. Other form of $R$ giving different neutrino interaction pattern might improve the sensitivity.

\subsection{Indirect Detection}
\label{sec:pheno_ID}

As we have seen in~\autoref{sec:pheno_DD}, although the perspectives of probing our model through direct detection are slightly better than in a scenario where the freeze-in production occurs in the usual radiation-dominated Universe, it is still hard to find it in the near future (see~\autoref{fig:directdetection}). The main reason for this is that the DM interaction strength with the SM (via the heavy neutrino) is still very small, suppressed by $m_\nu/m_N$ as well as by loop, which hampers the DM detection in direct experiments.

Nevertheless, space and ground-based telescopes can place stringent bounds on the dark matter annihilation cross-section and lifetime (see~\cite{Gaskins:2016cha} for a recent review). In our case, if dark matter is heavier than the heavy neutrino ($m_{DM}> m_{N}$), it may annihilate into heavy neutrinos that further decay into SM particles ($\nu h$, $\nu Z$, $l^\pm W^\mp$, $l^\pm W^\mp \gamma$, $\bar \nu \nu \nu$, $\nu l^+ l^-$) ~\cite{Pospelov:2007mp,Tang:2015coo,Campos:2017odj,Yang:2020vxl}, generating neutral and charged cosmic rays after hadronization and parton showers. In particular, over the past years, excesses of gamma-rays has been observed by several experiments, including {\tt INTEGRAL/SPI}~\cite{Jean:2005af}, {\tt Fermi-LAT}~\cite{TheFermi-LAT:2017vmf} and {\tt H.E.S.S.}~\cite{Abdallah:2016ygi}. This excess of high-energy photons can be sourced by DM annihilation, and, therefore, it can be a possible DM signature. An interesting feature of the neutrino portal DM is that, although a gamma-ray line signature is not expected, a distinct continuum gamma-ray signal might be present, as studied in~\cite{Tang:2015coo,Batell:2017rol} for the thermal DM case. In this section, we present the annihilation cross-section of DM into heavy neutrinos and show the constraints placed on it by {\tt Fermi-LAT} and {\tt H.E.S.S.}. 

In the non-relativistic regime ($s \approx 4 m_{\chi/S}^2(1+v^2/4)$, with $v$ the relative velocity between DM particles), the leading order of the annihilation cross-sections for the processes $\bar{\chi} \chi \to N N$ and $S^*S\to N N$ are given by
\begin{align}
    \sigma_{\bar \chi \chi \to N N} v \Big|_{v\approx 0} =& \frac{|\lam|^4}{16\pi} \sqrt{1-\frac{m_N^2}{m_\chi^2}}\frac{2m_\chi^2-m_N^2}{(m_\chi^2+m_S^2-m_N^2)^2}\,, \\
    \sigma_{S^*S \to N N} v \Big|_{v\approx 0} = & \frac{|\lam|^4}{8\pi}\left(1-\frac{m_N^2}{m_S^2}\right)^{3/2}\frac{m_N^2}{(m_\chi^2+m_S^2-m_N^2)^2}\,.
\end{align}

In Ref.~\cite{Campos:2017odj}, the authors constrain a generic dark matter annihilation cross-section into a pair of right-handed neutrinos in the context of the type-I seesaw, based on the amount of gamma-rays that can be produced via their two and three body decays. The constraints come from the {\tt Fermi-LAT} and {\tt H.E.S.S.} observations of Milky Way's dwarf spheroidal galaxies and center, respectively. They have considered one-flavor right-handed neutrinos separately with masses between $10$\,GeV and $1$\,TeV. For $m_\chi \sim 10$\,GeV, they have found $\la \sigma v \ra \lesssim 4 \times 10^{-27}\text{cm}^3/\text{s}$ ({\tt Fermi-LAT}), while the most stringent constraint for heavier dark matter comes from {\tt H.E.S.S.}, with $\la \sigma v \ra \lesssim 4 \times 10^{-26}\text{cm}^3/\text{s}$ for $m_\chi \sim 2$ TeV.

\begin{figure}[!t]
    \centering
    \includegraphics[width=\textwidth]{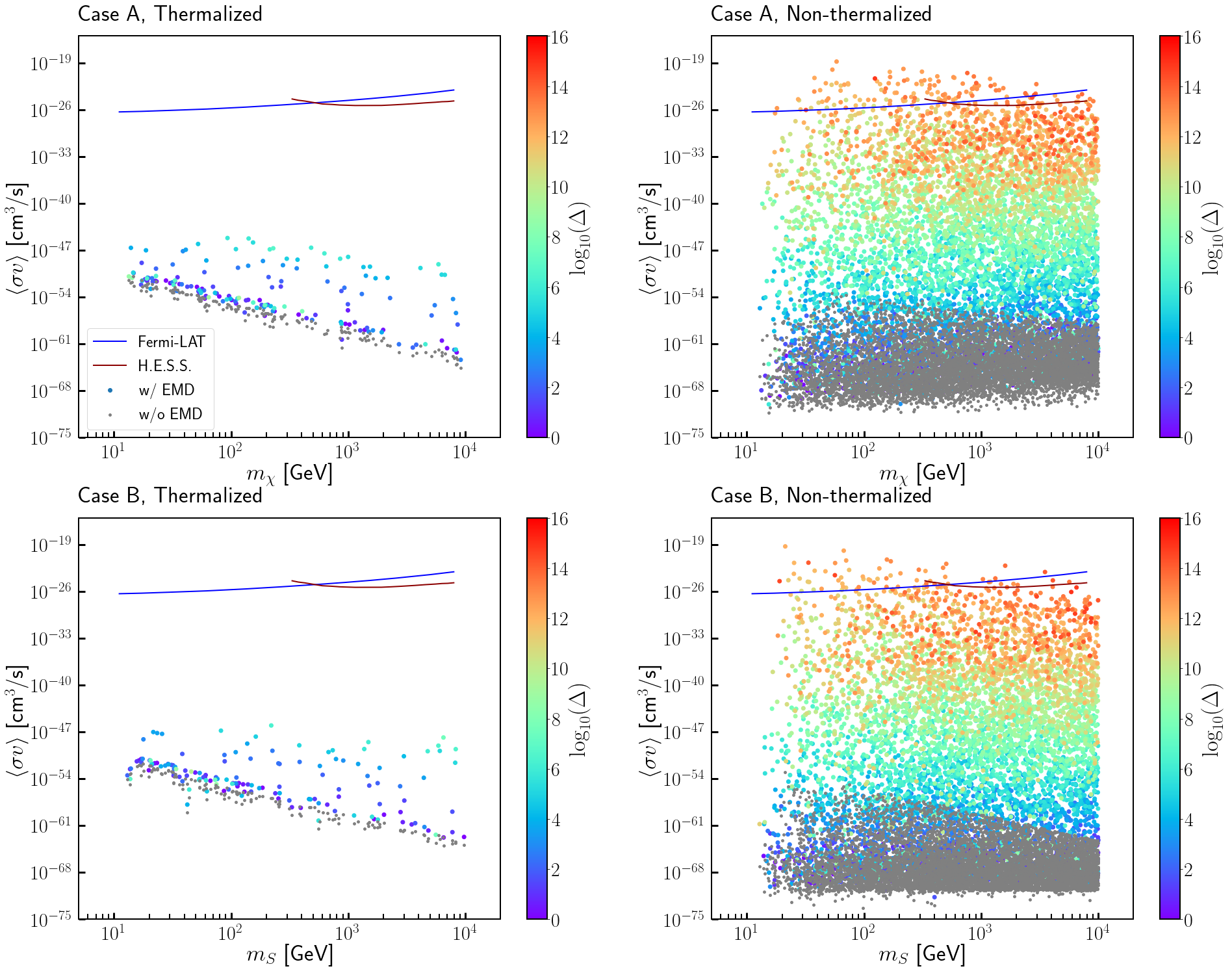}
    \caption{
        The annihilation cross section of the dark matter into heavy neutrino with (color points) and without (grey points) EMD for both Case A (upper panels) and Case B (lower panels). The heavy neutrino is considered to be thermalized (left panels) and non-thermalized (right panels) respectively. The blue and red lines indicate the constraints from {\tt Fermi-LAT} and {\tt H.E.S.S.} respectively~\cite{Campos:2017odj}.
    }
    \label{fig:indirectdetection}
\end{figure}

In~\autoref{fig:indirectdetection}, we show the annihilation cross section into heavy neutrinos pairs versus dark matter masses for both Case A and Case B with (colored points) and without (grey points) EMDE with both $N$ thermalized (left panels) and non-thermalized (right panels). Similar to the direct detection, with EMDE, larger $\lambda_\chi$ is needed to compensate the faster expansion and then to achieve the observed dark matter relic abundance. The largest annihilation cross section that can be achieved with EMDE is roughly $10^{-19} \text{cm}^3/\text{s}$ for both Case A and Case B, with $N$ non-thermalized. As pointed out already in~\autoref{sec:pheno_DD}, the case where heavy neutrinos are non-thermalized with the cosmic bath provides enhanced relevant cross-sections. This happens because the most efficient DM production mode is the exchange of $N$ via the s-channel, which is suppressed by Yukwawa couplings and, therefore, the coupling between $N$ and the DM candidate must be larger to attain the observed DM abundance. The limits from {\tt Fermi-LAT} (blue line) and {\tt H.E.S.S.} (red line)~\cite{Campos:2017odj} for $m_N = 10$ GeV are also shown in~\autoref{fig:indirectdetection} as a guidance. Notice, though, that since all points in this plot are for $m_N<m_\chi$, we expect comparable bounds on our parameter space for different values of $m_N$. Also, as we discussed in~\autoref{sec:thermalization_N}, when $N$ is much lighter than $S/\chi$, the thermalization of $N$ is more difficult. Thus, there are fewer points in the left panels of~\autoref{fig:indirectdetection} than in the right panels. We come therefore to our most important result: reasonably sizable FIMP couplings, allowed by long enough EMDE scenarios which, in turn, avoid the thermalization of heavy neutrinos, can already be tested by the current indirect detection experiments.

Future searches for gamma-ray and cosmic-ray signals with existing ({\tt H.E.S.S.-II}~\cite{Rinchiuso:2019rrh}) and planned ({\tt CTA}~\cite{Silverwood:2014yza}, {\tt GAMMA-400}~\cite{GALPER2013297}) experiments might become able to probe DM annihilations cross-sections into SM particles as low as order $10^{-28} \text{cm}^3/s$ in the light dark matter mass regime~\cite{Bringmann:2012ez,Ackermann:2012qk,Bergstrom:2012vd,Toma:2013bka,Garny:2013ama,Giacchino:2013bta,Pierre:2014tra,Clark:2019nby}. Indeed, potential gamma-ray lines from DM annihilation might be finally confirmed or ruled-out~\cite{Bergstrom:2012vd}. Also, the constraints on DM annihilation strongly depend on the halo profile which is subject to uncertainties. In particular, the existing limit on DM annihilation in the Galactic Center can be as low as $10^{-28} \text{cm}^3/s$ for contracted DM profiles~\cite{Hooper:2012sr}.

Interestingly, in the context of an early matter-dominated era, matter perturbations start growing linearly with the scale factor, which in principle have testable consequences~\cite{Easther:2010mr,Barenboim:2013gya,Fan:2014zua,Zhang:2015era}. Even a short period of linear growth could allow for perturbation modes to enter the nonlinear regime during the radiation era, leading to the early formation of very small and dense microhalos of DM particles\footnote{An upper bound on the duration of the EMDE could be invoked in order to avoid the formation of microhalos before reheating. They would be dominated by the matter content driving the EMDE and could be destroyed after it decays completely~\cite{Blanco:2019eij}.}~\cite{Gelmini_2008,Erickcek:2015jza,Erickcek:2015bda,Blanco:2019eij}. A long enough EMDE could also lead to the generation of detectable gravitational waves~\cite{Assadullahi:2009nf,DEramo:2019tit}. 

The presence of the EMDE-induced microhalos nowadays would boost the DM annihilation rate relevant for indirect detection searches by many orders of magnitude, depending on the duration of the EMDE and DM free-streaming length. The gamma-ray signal that could come from DM annihilation inside the microhalos is similar to the signal from DM decay, as it depends on the dark matter density instead of its square~\cite{Blanco:2019eij}. As a consequence, the \textit{current} lower bounds on dark matter lifetime might be translated into upper bounds on dark matter annihilation cross-sections. Considering the {\tt Fermi-LAT} observations of the isotropic gamma-ray background, the authors of Ref.~\cite{Delos:2019dyh} set upper bounds on dark matter annihilation into $b\bar b$ inside unresolved microhalos of the order $\la \sigma v \ra \lesssim 10^{-32} \text{cm}^3/\text{s}$, for $m_\dm \sim 10$ GeV. They have also shown that the {\tt Fermi-LAT} constraints on DM annihilations within the Draco dwarf galaxy are complementary to their conservative bounds, $\la \sigma v \ra \lesssim 4 \times 10^{-31} \text{cm}^3/\text{s}$, and could be used to distinguish them from dark matter decays, in the case of a future discovery.

\section{Conclusions}
\label{sec:conclusions}

In this work, we have studied the freeze-in production of FIMP dark matter through the neutrino portal with a modified cosmological history. In terms of the particle physics content of the model, we considered a scenario where, in addition to the SM sector, there is a hidden sector comprised of a fermion, $\chi$, and a complex scalar, $S$, with the lightest one being the DM candidate. The mass of the SM neutrinos is generated by the Type-I seesaw mechanism and we consider three degenerated heavy neutrinos, $N$, mediating the interactions between the SM and the hidden sectors. We explored the possibility of the heavy neutrinos being in thermal equilibrium with the cosmic bath, as well as the case where they are non-thermalized, to assess the impact of these hypothesis on the model. In regards to cosmology, we assumed that an EMDE took place and dominated the evolution of the Universe for some period between inflation and BBN, making it to expand faster. Then, this matter component decayed into SM degrees of freedom, reheating the visible sector, and the usual radiation-dominated phase took over. After characterizing the evolution of the Universe with an EMDE for a generic FIMP model, we computed the present DM abundance through the neutrino portal, via the freeze-in mechanism, evaluated the conditions for the freeze-in regime to hold and checked the thermalization condition of the heavy neutrinos. 

We found that an EMDE requires larger FIMP couplings to achieve the observed DM relic abundance. The reason is twofold: 1) to overcome a faster expansion, as during an EMDE, as well as an entropy production period, particles need to interact faster; 2) to overcome the dilution due to an EMDE with an initial DM overproduction.

The most important consequence of having larger couplings in the context of a FIMP model is the possibility of making it testable. In this work, we investigated the detectability of our scenario in direct and indirect detection experiments. We have found that an EMDE is able to significantly enhance the cross-sections relevant for both direct and indirect searches. Moreover, when $N$ is not part of the thermal bath, the only efficient production mode of FIMPs is via the s-channel exchange of $N$ in Higgs-lepton annihilations. Since it is suppressed by the Yukawa couplings, the couplings between $N$ and the dark matter candidates need to be larger in order to agree with the relic density constraints while still respecting the freeze-in regime. Therefore, non-thermalized heavy neutrinos manage to enhance even more the relevant cross-sections.

Regarding direct detection, the loop-induced nuclear recoil due to the fermionic FIMP is always much below the current bounds imposed by {\tt XENON-1T}, as well as the prospected ones from {\tt XENON-nT}. 

As expected for a neutrino portal model, the indirect signals from DM annihilation in dense regions is a better way of probe. In this case, with the larger couplings allowed by an EMDE, our FIMP candidates could efficiently annihilate into heavy neutrinos which then decay and produce cosmic rays. We have shown that a long enough EMDE makes possible to bring the FIMP annihilation cross-section into heavy neutrinos to the current sensitivity of {\tt Fermi-LAT} and {\tt H.E.S.S.} experiments. Hence, in the context of an EMDE, motivated by many extensions of the standard model of particle physics, indirect detection experiments can already test FIMPs and, in the case of a detection, it could also provide some hint about the evolution of the Universe.

In summary, the proposed model offers a viable DM candidate whose experimental signatures can be already tested through indirect detection experiments, showing that freeze-in DM can be probed at present. We should stress that both the EMDE and the non-thermalization of the heavy neutrinos play a key role on our model: in the case where the heavy neutrinos are not in thermal equilibrium with the cosmic plasma and the Universe undergoes an early-matter dominated phase, larger DM couplings are required to attain the observed DM abundance, which translates into a richer phenomenology.

\acknowledgments 

We would like to thank Heather Logan, Clarissa Siqueira, Tommi Tenkanen, Alberto Tonero and Ville Vaskonen for enlightening discussions. C.C. and M.D are supported by the Arthur B. McDonald Canadian Astroparticle Physics Reasearch Institute. T.M. is supported by the Israel Science Foundation (Grant No. 751/19), the United States-Israel Binational Science Foundation (BSF) (NSF-BSF program Grant No. 2018683) and the Azrieli foundation. Y. W is supported by the Natural Sciences and Engineering Research Council of Canada (NSERC). L. Y. is supported by the National Key Research and Development Program of China (Grant No. 2017YFA0402201).

\appendix
\section{Squared Amplitudes and Decay Widths}
\label{sec:amplitudes}

Here we report the non-averaged squared amplitudes and the decay widths relevant for the freeze-in process, with the Feynman rules following the conventions of Ref.~\cite{Denner:1992vza}.

The squared amplitude for a process $N^j \to \bar \psi_i \phi$, where $N^j$ is the heavy neutrino, $\psi_i$ is a Dirac or Majorana fermion and $\phi$ is a real scalar, with interaction strength $\lambda_{N^j\psi \phi}$. is given by:
\begin{equation}
|\M|^2_{\bar N_R^j \to \bar \psi_i \phi} = |\lambda_{N_R^j\psi \phi}|^2 \left(m_N^2 + m_\psi^2 - m_\phi^2\right)\,.
\end{equation}
The total decay width of $N^j$ is a function of temperature due to the thermal mass of the Higgs field, $m_{H^0}(T) = m_{H^+}(T) \approx c_h T$. It is given by:
\begin{equation}
\begin{split}
\Gamma_{N^j} &= 2 \Gamma_{N^j \to \bar \chi S} + 2 \Gamma_{N^j \to \nu_i H^0} + 2 \Gamma_{N^j \to l_i H^+}\\ 
&= \frac{m_N}{16 \pi} \Big[ |\lam^j|^2 (1+r_\chi^2-r_S^2) \sqrt{\lambda(1,r_\chi^2,r_S^2)} + \\
&\hspace{1.5cm} + |Y_\nu^{ij}|^2 \left(1+r_{\nu_i}^2 - \frac{m_{H^0}^2(T)}{m_N^2} \right) \sqrt{\lambda\left(1,r_{\nu_i}^2,\frac{m_{H^0}^2(T)}{m_N^2}\right)} \\
&\hspace{1.5cm} + |Y_\nu^{ij}|^2 \left(1+r_{l_i}^2 - \frac{m_{H^+}^2(T)}{m_N^2} \right) \sqrt{\lambda\left(1,r_{l_i}^2,\frac{m_{H^+}^2(T)}{m_N^2}\right)},
\end{split}
\label{NRdecay}
\end{equation}
recalling that $\lambda(x,y,z) = (x-(\sqrt{y}+\sqrt{z})^2)(x-(\sqrt{y}-\sqrt{z})^2)$ is the K\"allen function and $r_i \equiv m_i/m_N$.

For the s-channel process, the following squared amplitudes contribute:
\begin{equation}
|\M|^2_{\nu_i H^0 \to \chi S^*} = 2 |\lam|^2 |Y_\nu^{ij}|^2 \frac{m_N^2(m_\nu^2+m_\chi^2-t)}{(s-m_N^2)^2+m_N^2\Gamma_N^2},
\end{equation}
and
\begin{equation}
|\M|^2_{\nu_i H^0 \to \bar \chi S} 
= 2 |\lam|^2 |Y_\nu^{ij}|^2 \frac{ s(s+t) + m_\nu^2 m_\chi^2 - m_S^2(s+m_\nu) - m_{H^0}^2 (s+m_\chi^2-m_S^2) }{(s-m_N^2)^2+m_N^2\Gamma_N^2}.
\end{equation}
The factors of $2$ account for the contribution of the anti-particles.

Regarding the t-channels contributing for the production of $\chi$ and $S$, we have: 
\begin{equation}
\begin{split}  
|\M|^2_{N N \to \bar \chi \chi} 
= |\lam|^4 \Big[ &\frac{(m_N^2+m_\chi^2-t)^2}{(m_S^2-t)^2}+\frac{(m_N^2+m_\chi^2-s-t)^2}{(s+t+m_S^2-2(m_N^2+m_\chi^2))^2}\\ 
& - \frac{2m_N^2(s-2m_\chi^2)}{(m_S^2-t)(s+t+m_S^2-2(m_N^2+m_\chi^2)} \Big]
\end{split},
\end{equation}
and
\begin{equation}
\begin{split}  
|\M|^2_{N N \to S^* S} 
= |\lam|^4 \Big[ &\frac{-t(s+t)+2m_S^2t-(m_N^2-m_S^2)^2}{(m_\chi^2-t)^2}\\
& + \frac{-t(s+t)+2m_S^2t-(m_N^2-m_S^2)^2+2m_N^2(s+2t-2m_N^2)}{(s+t+m_\chi^2-2m_N^2-2m_\chi^2)^2}\\ 
& - \frac{4m_N^2(m_N^2-m_S^2)}{(m_\chi^2-t)(s+t+m_\chi^2-2m_N^2-2m_S^2)} \Big] \,.
\end{split}
\end{equation}

\section{Evolution of a Matter-radiation System}
\label{sec:freezein_evolution}

Here we give more details on the solution of the coupled set of Boltzmann fluid equations~\autoref{equ:energydensities} and~\autoref{equ:entropy}, shown in~\autoref{Fig:thermalhistory}.
 
From $aH(a)=da/dt$, valid throughout all the expansion, we can translate the evolution over time to the evolution over the scale factor. We re-scale the energy density of the matter component by $\Phi \equiv \rho_M/T_r a^3$. We also define the comoving amount of radiation by ${\cal N}_R = \rho_R a^4$, and use the dimensionless variable $A\equiv a T_r$ as evolution parameter. Notice that ${\cal N}_M = A \Phi$. We therefore need to solve:
\begin{equation}
\label{set}
\begin{cases}
\frac{d \Phi}{dA} = - c_1 \frac{A \Phi}{\sqrt{A \Phi + {\cal N}_R}}\\
\frac{d {\cal N}_R}{dA} \approx c_1 B_R \frac{A^2 \Phi}{\sqrt{A \Phi + {\cal N}_R}}\,,
\end{cases}
\end{equation}
where we define the recurrent constant $c_1 = \sqrt{3}\kappa\frac{\pi\sqrt{g_e(A_r)}}{3\sqrt{10}}$ and assume that the production of dark matter from radiation does not change $\rho_R$ significantly. 

The evolution of the entropy reads:
\begin{equation}
\label{Sevol}
\frac{dS}{dA} = c_1 B_R  \left(\frac{2\pi^2}{45} g_\mathfrak{s}(A)\right)^{1/3}
\frac{A^2 \Phi S^{-1/3}}{\sqrt{A \Phi + {\cal N}_R}}\,. 
\end{equation}

In the case of a Universe filled with radiation and matter, the Hubble rate is explicitly:
\begin{equation}\label{Eq:HubbleA}
H(A) = \frac{T_r^2}{\sqrt{3}M_P} A^{-2}\sqrt{A \Phi + {\cal N}_R}\,,
\end{equation}
so that we have $H(a) \propto a^{-2}$ when a radiation content dominates and $H(a) \propto a^{-3/2}$ when a matter content dominates, as we can see in~\autoref{Fig:thermalhistory}. 

In this work, we assume that the initial conditions for the EMDE, $S(T_i)$ and ${\cal N}_R(T_i)$ are set by a given inflationary model. These values are found after solving the above set of equations for the inflaton-radiation system. From~\autoref{Eq:HubbleA}, we see that the initial condition for $\Phi$ is given by $\Phi_I/A_I^3 = 3M_P^2 H_I^2/T_r^4$. The current observational constraints on the post-inflationary reheat temperature poses $T_\rh \lesssim 7 \times 10^{15}$ GeV~\cite{rehagen_dark_2015}. In order to explore the scenario in which there was a large entropy production prior BBN, we are interested in the case where $T_\rh > T_i \gg T_r$. As a benchmark value, we consider $T_\rh = 7 \times 10^{15}$ GeV and the Hubble rate after inflation $H_I = 3.3 \times 10^{14}$ GeV~\cite{Akrami:2018odb}, with $A_I = 1$. We find $T_\max \simeq 7.9 \times 10^{15}$ GeV, $S_i \equiv S (T_i) = 897$ and ${\cal N}_R^i \equiv {\cal N}_R (T_i) = 1790$.

It is interesting to notice that the initial conditions fix the amount of entropy which is going to be produced from $T_e$ to $T_r$:
\begin{equation}\label{Sinicond}
\frac{S_r}{S_i} \sim \Delta = \frac{m_M Y_M (T_i)}{T_r} = \frac{m_M}{T_r}\frac{N_M (T_i)}{S_i} = \frac{\Phi_i}{S_i} \,,
\end{equation}
where we have used $\Phi = N_M m_M/T_r$, in terms of the total number $N_M = n_M a^3$.

Choosing $\Delta$ as a free parameter, we can therefore find the initial condition $\Phi_i \equiv \Phi(T_i)$, which sets the moment of ERD-EMDE equality (${\cal N}_R^i = {\cal N}_M^i$), $A_i = {\cal N}_R^i/\Phi_i$. 

Let us now find out when the entropy would start to be produced, from the initial conditions. Before the decay of $\m$, while $\Phi \approx \Phi_i$, we can solve~\autoref{set} for ${\cal N}_R$ from $A_e$ to a given $A$:
\begin{equation}
\label{comoving rad}
{\cal N}_{R}\left(A\right) \approx {\cal N}_{R}(A_{e})+\frac{2}{5}c_{1}B{}_{R}\sqrt{\Phi_i}(A^{5/2}-A_{e}^{5/2}).
\end{equation}

As long as we have a thermal bath, we can define temperature from the energy density of radiation:
\begin{equation}\label{TA}
\begin{split}
T(A) &\equiv \left(\frac{\pi^2}{30}g_e(A)\right)^{-1/4} \frac{T_r}{A} {\cal N}_R (A)^{1/4} \\
& = \left(\frac{\pi^2}{30}g_e(A)\right)^{-1/4} \frac{T_r}{A} \left( {\cal N}_R (A_e) + \frac{2}{5} c_1 B_R \sqrt{\Phi_i} (A^{5/2} - A_e^{5/2})  \right)^{1/4}.
\end{split}
\end{equation}

By equaling both terms of the equation above we can find when the entropy will start to be produced:
\begin{equation}
    A_e \approx \left(\frac{5}{2}\frac{{\cal N}_R^i}{c_1 B_R \sqrt{\Phi_i}}\right)^{2/5} \,.
\end{equation}

Generalizing the approach used in Ref.~\cite{Giudice:2000ex}, the temperature of the thermal bath during an EP period goes as $A^{-3/8}$ (see~\autoref{Fig:thermalhistory}) and is approximately given by:
\begin{equation}\label{Tep}
T(A) = T_e k(g_e(A)) \left(\left(\frac{A}{A_e}\right)^{-3/2} - \left(\frac{A}{A_e}\right)^{-4} \right)^{1/4} \approx T_e k(g_e(A)) (A/A_e)^{-3/8} \,,
\end{equation}
where the function $k(A)$ is defined such that $T(A)=T_e$ at the maximal point:
\begin{equation}
k(g_e(A)) = \left(\frac{8^8}{3^3 5^5}\right)^{1/20} \left(\frac{g_e(A_e)}{g_e(A)}\right)^{1/4}  \,. 
\end{equation}

As expected from~\autoref{equ:EMDE_relations}, the hierarchy between $T_e$ and $T_r$ also depends on the initial condition for the matter content:
\begin{equation}\label{Tedef}
\frac{T_e}{T_r} = \left(\kappa B_R\frac{3^{11/10} 5^{1/2}}{2^{23/10}\pi \sqrt{g_e(T_r)\frac{g_e(T)}{g_e^2(T_e)}}}\right)^{1/4} \left(\frac{\Phi (T_i)}{A_e^3}\right)^{1/8}\,. 
\end{equation}

Finally, taking as free parameters $T_\rh = 7 \times 10^{15}$ GeV, $H_I = 3.3 \times 10^{14}$ GeV, $T_r = 4$ MeV and $T_i = 10^{14}$GeV, we find $T_\max = 7.9 \times 10^{15}$ GeV, $T_e = 7.6$ GeV and $\Delta = 1.9 \times 10^{16}$. In this case, we have $A_i = 1.1 \times 10^{-16}, A_e = 1.4 \times 10^{-3}$ and $A_r = 7.8 \times 10^{5}$.

\bibliographystyle{JHEP}
\bibliography{FIfermionportal}

\end{document}